\documentclass{elsart}
\usepackage{latexsym}
\usepackage{amsmath}
\usepackage{mathrsfs}
\usepackage{amssymb}
\usepackage{fancybox}
\usepackage{mathbbol}
\usepackage{bm}
\usepackage{cmap}
\usepackage{fancyhdr}
\usepackage{subfigure}
\usepackage{graphicx}

\begin{document}

\begin{frontmatter}
\title{Intrinsic localized modes in coupled DNLS equations from the
  anti-continuum limit}
\author[usa]{K. Li}
\author[usa]{P.G. Kevrekidis}
\author[uk]{H. Susanto}
\author[gre]{V. Rothos}
\address[usa]{Department of mathematics and statistics, university of Massachusetts, Amherst, Massachusetts 01003-4515, USA}
\address[uk]{School of mathematical sciences, university of Nottingham, University Park, Nottingham, NG7 2RD, United Kingdom}
\address[gre]{Department of mathematics, physics and computational sciences,
faculty of engineering, Aristotle university of Thessaloniki, Thessaloniki 54124, Greece}

\begin{abstract}
In the present work, we generalize earlier considerations for
intrinsic localized modes consisting of a few excited sites, as developed
in the one-component discrete nonlinear Schr{\"o}dinger equation
model, to the case of two-component systems. We consider all the
different combinations of ``up'' (zero phase) and ``down'' ($\pi$
phase) site excitations and are able to compute not only the corresponding
existence curves, but also the eigenvalue dependence of the small
eigenvalues potentially responsible for instabilities, as a function
of the nonlinear parameters of the model representing the self/cross
phase modulation in optics and the scattering length ratios in the
case of matter waves in optical lattices. We corroborate these
analytical predictions by means of direct numerical computations.
We infer that all the modes
which bear two adjacent nodes with the same phase are unstable in
the two component case and the only solutions that may be linear
stable are ones where each set of adjacent nodes, in each component
is out of phase.
\end{abstract}

\end{frontmatter}

\section{Introduction}

The theme of nonlinear dynamical lattices and especially of
the prototypical equation of the Discrete Nonlinear Schr{\"o}dinger (DNLS)
type~\cite{pgk} has been a widely studied one over the last decade, perhaps
especially so due to its applications in both the field of nonlinear
optics, as well as in that of atomic physics.

The realization of this type of models in the context of matter waves
arose due to the examination of the dynamics of Bose-Einstein
condensates (BECs) in periodic, so-called optical lattice, potentials.
In that realm, the reduction of the prototypical mean-field model
of the Gross-Pitaevskii equation with a periodic potential, in the
superfluid regime, via a tight binding approximation, naturally yields
the DNLS as the canonical lattice model for the BEC
``droplets''~\cite{konotop,markus2,ourbook}.

On the other hand, the nonlinear optics realization in fabricated
AlGaAs waveguide arrays \cite{7} has been one that preceded the atomic
physics realization and yielded numerous insights about the interplay of
discrete lattice dynamics with the effects of nonlinearity, including
but not limited to Peierls-Nabarro barriers and discrete solitons,
modulational instabilities, 
gap solitons, diffraction and diffraction management; see e.g.
\cite{general_review1,general_review2,general_review3,review_opt,discrete_opt} for a number of relevant reviews.
Another area of optical applications which has seen considerable
development over the past decade is that of optically induced
lattices in photorefractive crystals such as
Strontium-Barium-Niobate~\cite{efrem,moti1,moti2}. Although the latter
setting is not directly modeled by the DNLS (chiefly due to the
photorefractive, saturable nature of the nonlinearity), it has enabled
the experimental observation of numerous solitary/localized patterns
that were predicted in the context of the DNLS;
see e.g. the reviews~\cite{discrete_opt,rev_moti1}.

A topic that has been of particular interest in both the
atomic and the optical applications has been that of multi-component
dynamics. Among the many recent experimental results related
to the interaction of two different frequencies or polarizations
in optics or of two hyperfine atomic states or two different atoms
in BEC, we highlight only a few. The first experimental observations
of discrete vector solitons in waveguide arrays were reported
in~\cite{christo},
while the emergence of multipole patterns in two-component settings
within photorefractive crystals was presented in~\cite{zhig_two}, and
more recently more complex patterns including even dark-bright lattice
solitons were reported experimentally~\cite{detlef}. In the BEC
case, experiments with mixtures of different
spin states of $^{87}$Rb \cite{myatt,us_dh} or $^{23}$Na \cite{stamper}
and even ones of different atomic species such as
$^{41}$K--$^{87}$Rb \cite{KRb} and $^{7}$Li--$^{133}$Cs
\cite{LiCs} have been realized and although localized modes
in optical lattices have not been probed in the form considered
herein, certainly such experiments appear to be within the current
experimental reach~\cite{morsch3}.

Our scope in the present communication is to explore systematically
the existence and stability properties of two-component localized
modes, motivated by the above applications. The most fundamental
excitations (single site) in the coupled DNLS equations model were
explored a considerable while ago not only in one- but also in higher
dimensions; see for a relevant discussion~\cite{hudock} and references
therein. In fact, in higher dimensions also vortex states were
considered~\cite{pelin}. However, in the one-dimensional realm,
we are not aware of a multi-component generalization of the theory
of~\cite{pelin1} (for the existence theory, see also~\cite{konotop1})
which enables the systematic quantification of
the existence and stability properties of localized modes consisting
of an arbitrary number of sites. Here, we present the theory of how
to address the problem of existence and stability in the two-component
setting, which paves the way also towards the generalization to a
higher
number of components. Aside from providing the general theory, we
consider numerous special case examples involving excitation of
two- and also of three-sites in both components with either zero phase
(an ``up'' site) or with $\pi$ phase (a ``down'' site). For all of
these configurations we examine their stability properties as a
function,
e.g., of their nonlinear coefficients (dubbed $g_{ij}$), which
stand for the self- and cross-phase modulation parameters in optics
and for the scattering lengths (or more accurately their ratios)
in the realm of BEC. Interestingly, we find that independently of
the nature of the second component and of the inter- and
intra-component
coupling strengths, should two adjacent sites in any one component
be of the same phase, the configuration is found to be unstable.
The only potentially stable configurations consist of nodes
with alternating adjacent phases, similarly to what is known for
the one-component problem~\cite{pelin1}.

Our presentation is structured as follows. Section II presents
the general existence and stability theory. Section III examines
special case examples of configurations. Finally, section IV
summarizes our findings and presents our conclusions, as well
as topics for potential further studies.

\section{General theory}

Our fundamental model will be that of the two coupled
DNLS equations (see e.g.~\cite{hudock,pelin1,interlaced}) of the form:
\begin{eqnarray}
\label{DNLS}
\left\{
\begin{array}{ll}
i\stackrel{\cdot}{U_n}+(g_{11}|U_n|^2+g_{12}|V_n|^2)U_n+\epsilon\Delta U_n=0\\
i\stackrel{\cdot}{V_n}+(g_{12}|U_n|^2+g_{22}|V_n|^2)V_n+\epsilon\Delta V_n=0.
\end{array}
\right.
\end{eqnarray}
The localized modes (often referred to also as ``discrete solitons'')
 are given by the time-periodic solutions of the DNLS equation
\begin{eqnarray}
\left\{
\begin{array}{ll}
U_n(t)=\phi_n e^{i(\mu-2\epsilon)t+i\xi_0} \\
V_n(t)=\psi_n e^{i(\mu-2\epsilon)t+i\zeta_0},
\end{array}
\right.
\end{eqnarray}
where $\mu\in\mathbb R,\ \phi_n,\psi_n\in\mathbb C,\ n\in\mathbb Z,\ \xi_0,\zeta_0\in\mathbb R$
is a parameter and $(\mu,\phi_n,\psi_n)$ solve the nonlinear difference equations on $n\in\mathbb Z$:
\begin{eqnarray}
\left\{
\begin{array}{ll}
(\mu-g_{11}|\phi_n|^2-g_{12}|\psi_n|^2)\phi_n=\epsilon(\phi_{n+1}+\phi_{n-1}) \\
(\mu-g_{12}|\phi_n|^2-g_{22}|\psi_n|^2)\psi_n=\epsilon(\psi_{n+1}+\psi_{n-1}) .
\end{array}
\right.
\end{eqnarray}
Notice that here we start by examining the simpler case where the
soliton
parameter $\mu$ (representing the chemical potential in the atomic
physics
case and the propagation constant in the optical case) is identical
in the two components. Nevertheless, this is not necessarily the case
given the nonlinear nature of the coupling and the presence of
an invariance under phase transformations in each of the involved
components. We will comment on the case with distinct values of
$\mu$ at the final section of this presentation.

Our principal existence result can be formulated as follows (the proof
follows a line directly analogous to that of~\cite{pelin1} for the
two-component case, imposing the relevant solvability conditions
that in turn lock the adjacent site phase differences to the
$0$, $\pi$ relative phases discussed below)

Assume that $\mu=1$ and $(g_{12}-g_{22}) (g_{12}^2-g_{11} g_{22})>0$, as well
as $(g_{12}-g_{11}) (g_{12}^2-g_{11} g_{22})>0$.
There exists $\epsilon_0>0$ such that the solution $\phi_n$ and $\psi_n$, $n\in\mathbb Z$,
are represented by the convergent power series for $0\le\epsilon<\epsilon_0$:
\begin{equation}
\label{seriesphi}
\phi_n=\phi_n^{(0)}+\sum_{k=1}^{\infty}{\epsilon^k\phi_n^{(k)}}
\end{equation}
and
\begin{equation}
\label{seriespsi}
\psi_n=\psi_n^{(0)}+\sum_{k=1}^{\infty}{\epsilon^k\psi_n^{(k)}},
\end{equation}
where
\begin{eqnarray}
\left(
\begin{array}{c}
\displaystyle\phi_n^{(0)} \\
\displaystyle\psi_n^{(0)}
\end{array}
\right)=
\left(
\begin{array}{c}
\displaystyle\sqrt{\frac{g_{12}-g_{22}}{g_{12}^2-g_{11}g_{22}}}e^{i\xi_n} \\
\displaystyle\sqrt{\frac{g_{12}-g_{11}}{g_{12}^2-g_{11}g_{22}}}e^{i\zeta_n}
\end{array}
\right),\
\end{eqnarray}
or
\begin{eqnarray}
\left(
\begin{array}{c}
\displaystyle\phi_n^{(0)} \\
\displaystyle\psi_n^{(0)}
\end{array}
\right)=
\left(
\begin{array}{c}
\displaystyle\frac{1}{\sqrt{g_{11}}}e^{i\xi_n} \\
0
\end{array}
\right),\
\end{eqnarray}
or
\begin{eqnarray}
\left(
\begin{array}{c}
\displaystyle\phi_n^{(0)} \\
\displaystyle\psi_n^{(0)}
\end{array}
\right)=
\left(
\begin{array}{c}
0 \\
\displaystyle\frac{1}{\sqrt{g_{22}}}e^{i\zeta_n}
\end{array}
\right),
\end{eqnarray}
with $\xi_n,\ \zeta_n\in\{0,\pi\}$, is the solution for
\begin{eqnarray}
\left\{
\begin{array}{ll}
(1-g_{11}|\phi_n|^2-g_{12}|\psi_n|^2)\phi_n=0 \\
(1-g_{12}|\phi_n|^2-g_{22}|\psi_n|^2)\psi_n=0.
\end{array}
\right.
\label{linsys}
\end{eqnarray}
for $n\in S=S_\phi=S_\psi=\{1,2,\cdots,N\}$ which is a finite set of
(excited) nodes of the lattice, and $\phi^{(0)}_n=\psi^{(0)}_n=0$ otherwise.

The above analytical expressions fully describe the
anti-continuum limit of uncoupled sites (i.e., $\epsilon=0$). In
particular,
they correspond to the cases of both fields being excited at each
lattice node, or of those of one field being excited only thereon.
The latter case was considered previously in~\cite{interlaced}, hence
in the present work, we will focus on genuine two-component
excitations. Notice also that in the above considerations (and in what
follows), we have set, without loss of generality, $\mu=1$.
Lastly, we should point out that for the two-excited-component
solutions to be meaningful, it should be the case that
$(g_{12}-g_{22}) (g_{12}^2-g_{11} g_{22})>0$ and similarly
that $(g_{12}-g_{11}) (g_{12}^2-g_{11} g_{22})>0$.
This condition stems from the necessity that the solutions of
Eq.~(\ref{linsys}) for $|\phi_n|^2$ and $|\psi_n|^2$ be positive.
It should
be added, however, that these inequalities should not necessarily be thought of
as binding either since if the opposite inequality signs hold, then
one can simply consider the case where $\mu$ has opposite sign.

The spectral stability of discrete solitons is studied with the standard linearization
\begin{eqnarray}
\left\{
\begin{array}{ll}
u_n(t)=e^{i(1-2\epsilon)t+i\xi_0}(\phi_n+a_ne^{\lambda t}+\overline{b}_ne^{\overline{\lambda}t})\\
v_n(t)=e^{i(1-2\epsilon)t+i\zeta_0}(\psi_n+c_ne^{\lambda t}+\overline{d}_ne^{\overline{\lambda}t}),
\end{array}
\right.
\end{eqnarray}
where $\lambda,a_n,b_n,c_n,d_n$ solve the linear eigenvalue problem on $n\in \mathbb Z$:
\begin{eqnarray}
\label{linearization}
\begin{array}{ll}
\alpha_n a_n-g_{11}\phi_n^2b_n-g_{12}\phi_n\psi_n(c_n+d_n)-\epsilon(a_{n+1}+a_{n-1})=i\lambda a_n\\
\alpha_n b_n-g_{11}\phi_n^2a_n-g_{12}\phi_n\psi_n(c_n+d_n)-\epsilon(b_{n+1}+b_{n-1})=-i\lambda b_n\\
\beta_n c_n-g_{22}\psi_n^2c_n-g_{12}\phi_n\psi_n(a_n+b_n)-\epsilon(c_{n+1}+c_{n-1})=i\lambda c_n\\
\beta_n d_n-g_{22}\psi_n^2d_n-g_{12}\phi_n\psi_n(a_n+b_n)-\epsilon(d_{n+1}+d_{n-1})=-i\lambda d_n,
\end{array}
\end{eqnarray}
where $\alpha_n=1-2g_{11}\phi_n^2-g_{12}\psi_n^2,\ \beta_n=1-2g_{22}\psi_n^2-g_{12}\phi_n^2$.

Let $\Omega=l^2(\mathbb Z,\mathbb C)$ be the Hilbert space of square-summable bi-infinite complex-valued
sequences $\{u_n\}_{n\in \mathbb Z}$, equipped with the inner product and norm
\begin{equation}
\label{innerproduct}
(\mathbf u,\mathbf v)_\Omega=\sum_{n\in \mathbb Z}\overline{u}_nv_n, \quad
\parallel\mathbf u\parallel_\Omega^2=\sum_{n\in \mathbb Z}|u_n|^2<\infty.
\end{equation}
We use bold notation $\mathbf u$ for an infinite-dimensional vector in $\Omega$ that consists of
components $u_n$ for all $n\in\mathbb Z$.\\
The stability problem (\ref{linearization}) is transformed through
the substitution
\begin{equation}
a_n=r_n+is_n, b_n=r_n-is_n, c_n=p_n+iq_n, d_n=p_n-iq_n,
\end{equation}
to the form
\begin{eqnarray}
\begin{array}{ll}
(1-3g_{11}\phi_n^2-g_{12}\psi_n^2)r_n-2g_{12}\phi_n\psi_np_n-\epsilon(r_{n+1}+r_{n-1})=-\lambda s_n\\
(1-g_{11}\phi_n^2-g_{12}\psi_n^2)s_n-\epsilon(s_{n+1}+s_{n-1})=\lambda r_n\\
(1-3g_{22}\psi_n^2-g_{12}\phi_n^2)p_n-2g_{12}\phi_n\psi_nr_n-\epsilon(p_{n+1}+p_{n-1})=-\lambda q_n\\
(1-g_{22}\psi_n^2-g_{12}\phi_n^2)q_n-\epsilon(q_{n+1}+q_{n-1})=\lambda p_n.
\end{array}
\end{eqnarray}

The matrix-vector form of the problem is then 
\begin{eqnarray}
\label{substitution}
\begin{array}{cc}
\mathcal L_+ \mathbf u=-\lambda\mathbf v\\
\mathcal L_-\mathbf v=\lambda\mathbf u,
\end{array}
\end{eqnarray}
where
\begin{equation}
\mathbf u=\left(\begin{array}{cc}
r_n\\p_n
\end{array}\right),
\mathbf v=\left(\begin{array}{cc}
s_n\\q_n
\end{array}\right),
\end{equation}
and $\mathcal L_\pm$ are infinite-dimensional symmetric matrices which have tridiagonal blocks as follows:
\begin{eqnarray}
\begin{array}{ccc}
(\mathcal L_+)_{n,n}=\left(\begin{array}{cc}
1-3g_{11}\phi_n^2-g_{12}\psi_n^2 & -2g_{12}\phi_n\psi_n\\
-2g_{12}\phi_n\psi_n & 1-3g_{22}\psi_n^2-g_{12}\phi_n^2
\end{array}\right)\\\\
(\mathcal L_-)_{n,n}=\left(\begin{array}{cc}
1-g_{11}\phi_n^2-g_{12}\psi_n^2 & 0\\
0 & 1-g_{22}\psi_n^2-g_{12}\phi_n^2
\end{array}\right)\\\\
(\mathcal L_\pm)_{n,n+1}=(\mathcal L_\pm)_{n+1,n}=\left(\begin{array}{cc}
-\epsilon & 0\\
0 & -\epsilon
\end{array}\right).
\end{array}
\end{eqnarray}

Equivalently, the stability problem (\ref{substitution}) is rewritten in the Hamiltonian form
\begin{equation}
\label{Hamiltonian}
\mathcal{JH}\bm\eta=\lambda\bm\eta,
\end{equation}
where $\bm\eta$ is the infinite-dimensional eigenvector, which consists of 4-blocks of $(r_n,p_n,s_n,q_n)^T$,
$\mathcal J$ is the infinite-dimensional skew-symmetric matrix, which consists of 4-by-4 blocks of
\begin{equation}
\mathcal J_{n,m}=\left(
\begin{array}{cccc}
0 & 0 & 1 & 0\\
0 & 0 & 0 & 1\\
-1 & 0 & 0 & 1\\
0 & -1 & 0 & 0
\end{array}\right)
\delta_{n,m},
\end{equation}
and $\mathcal H$ is the infinite-dimensional symmetric matrix, which consists of 2-by-2 blocks of
\begin{equation}
\label{H}
\mathcal H_{n,m}=\left(
\begin{array}{cc}
(\mathcal L_+)_{n,m} & 0\\
0 & (\mathcal L_-)_{n,m}
\end{array}\right).
\end{equation}
The representation (\ref{Hamiltonian}) follows from the Hamiltonian structure of the DNLS equation
(\ref{DNLS}), where $\mathcal J$ is the symplectic operator and $\mathcal H$ is the linearized Hamiltonian.
Using our above existence results,
the matrix $\mathcal H$ can be expanded (again for
$\epsilon$ small) into the power series
\begin{equation}
\mathcal H=\mathcal H^{(0)}+\sum_{k=1}^\infty\epsilon^k\mathcal H^{(k)},
\end{equation}
where
\begin{eqnarray}
\label{H0}
\begin{array}{cc}
\mathcal H_{n,n}^{(0)}=\left(
\begin{array}{cccc}
-2g_{11}{\phi_n^{(0)}}^2 & -2g_{12}\phi_n^{(0)}\psi_n^{(0)} & 0 & 0\\
-2g_{12}\phi_n^{(0)}\psi_n^{(0)} & -2g_{22}{\psi_n^{(0)}}^2 & 0 & 0\\
0 & 0 & 0 & 0\\
0 & 0 & 0 & 0
\end{array}\right), & n\in S,\\
\mathcal H_{n,n}^{(0)}=\left(
\begin{array}{cccc}
1 & 0 & 0 & 0\\
0 & 1 & 0 & 0\\
0 & 0 & 1 & 0\\
0 & 0 & 0 & 1
\end{array}\right), & n\in \mathbb Z\backslash S.
\end{array}
\label{H0n}
\end{eqnarray}

The solution of the difference equations is defined by the
power series (\ref{seriesphi}) and (\ref{seriespsi}),
and the pair $\phi_n^{(1)}, \psi_n^{(1)}$ solves the inhomogeneous problem
\begin{eqnarray}
\left\{
\begin{array}{ll}
(1-3g_{11}{\phi_n^{(0)}}^2-g_{12}{\psi_n^{(0)}}^2)\phi_n^{(1)}-2g_{12}\phi_n^{(0)}\psi_n^{(0)}\psi_n^{(1)}=\phi_{n+1}^{(0)}+\phi_{n-1}^{(0)}\\
(1-3g_{22}{\psi_n^{(0)}}^2-g_{12}{\phi_n^{(0)}}^2)\psi_n^{(1)}-2g_{12}\phi_n^{(0)}\psi_n^{(0)}\phi_n^{(1)}=\psi_{n+1}^{(0)}+\psi_{n-1}^{(0)}.
\end{array}
\right.
\end{eqnarray}
The symmetric matrix $\mathcal H$ is defined by (\ref{H}), where $\mathcal H^{(0)}$ is given by (\ref{H0})
and $\mathcal H^{(1)}$ consists of blocks:
\begin{equation}
\label{H1}
\mathcal H_{n,n}^{(1)}=\left(
\begin{array}{cc}
( {\mathcal L}_+^{(1)})_{n,n} & 0\\
0 & ({\mathcal L}_-^{(1)})_{n,n}
\end{array}\right),
\end{equation}
where
\begin{equation}
({\mathcal L}_+^{(1)})_{n,n}=\left(
\begin{array}{cc}
-6g_{11}\phi_n^{(0)}\phi_n^{(1)}-2g_{12}\psi_n^{(0)}\psi_n^{(1)} & -2g_{12}(\phi_n^{(0)}\psi_n^{(1)}+\psi_n^{(0)}\phi_n^{(1)})\\
-2g_{12}(\phi_n^{(0)}\psi_n^{(1)}+\psi_n^{(0)}\phi_n^{(1)}) & -2g_{12}\phi_n^{(0)}\phi_n^{(1)}-6g_{22}\psi_n^{(0)}\psi_n^{(1)}
\end{array}\right),
\end{equation}
\begin{equation}
({\mathcal L}_-^{(1)})_{n,n}=\left(
\begin{array}{cc}
-2g_{11}\phi_n^{(0)}\phi_n^{(1)}-2g_{12}\psi_n^{(0)}\psi_n^{(1)} & 0\\
0 & -2g_{22}\psi_n^{(0)}\psi_n^{(1)}-2g_{12}\phi_n^{(0)}\phi_n^{(1)}
\end{array}\right),
\label{L-1}
\end{equation}
\begin{equation}
\mathcal H_{n,n+1}^{(1)}=\mathcal H_{n+1,n}^{(1)}=\left(
\begin{array}{cccc}
-1 & 0 & 0 & 0\\
0 & -1 & 0 & 0\\
0 & 0 & -1 & 0\\
0 & 0 & 0 & -1
\end{array}\right),
\end{equation}
while all other blocks of $\mathcal H_{n,m}^{(1)}$ are zero.

Now, the non-self-adjoint eigenvalue problem
(\ref{substitution}) can be transformed to the self-adjoint problem
\begin{equation}
\mathcal M\mathbf u={\mathcal L}_+ {\mathcal L}_-\mathbf u=-\lambda^2\mathbf u.
\end{equation}

According to (\ref{H0n}),
\begin{equation}
({\mathcal L}_+^{(0)})_{n,n}=\left(
\begin{array}{cc}
-2g_{11}{\phi_n^{(0)}}^2 & -2g_{12}\phi_n^{(0)}\psi_n^{(0)}\\
-2g_{12}\phi_n^{(0)}\psi_n^{(0)} & -2g_{22}{\psi_n^{(0)}}^2
\end{array}\right),
\label{L+0}
\end{equation}
\begin{equation}
({\mathcal L}_-^{(0)})_{n,n}=\left(
\begin{array}{cc}
0 & 0\\
0 & 0
\end{array}\right).
\end{equation}

Therefore,
\begin{equation}
\mathcal M^{(1)}={\mathcal L}_+^{(0)} {\mathcal L}_-^{(1)},
\label{M1}
\end{equation}
where $\mathcal L_+^{(0)}$ is given by (\ref{L+0}) and $\mathcal
L_-^{(1)}$ is given by (\ref{L-1}), while the expansion
$\mathcal{M}=\epsilon\mathcal{M}^{(1)}+O(\epsilon^2)$ has been used.
This expansion is a natural generalization of the one-component
results of~\cite{pelin1}. As a result, if we identify the
eigenvalues $\gamma$ of $\mathcal M^{(1)}$, we can then trace the
eigenvalues of the full Hamiltonian problem according to
$\lambda^2 =-\epsilon \gamma$.

\begin{figure}[htp]
\centering
\scalebox{0.3}{\includegraphics{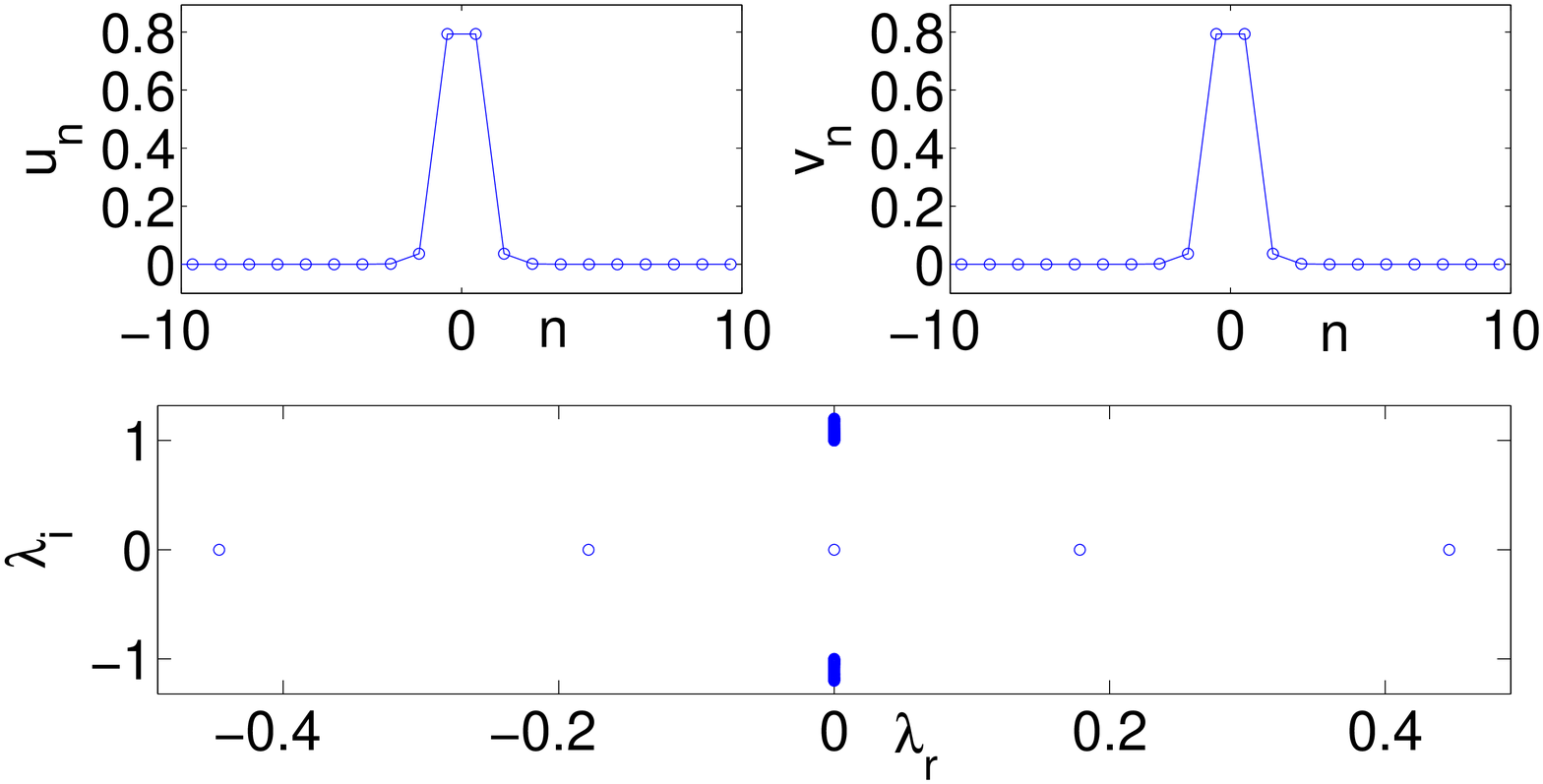}}
\scalebox{0.3}{\includegraphics{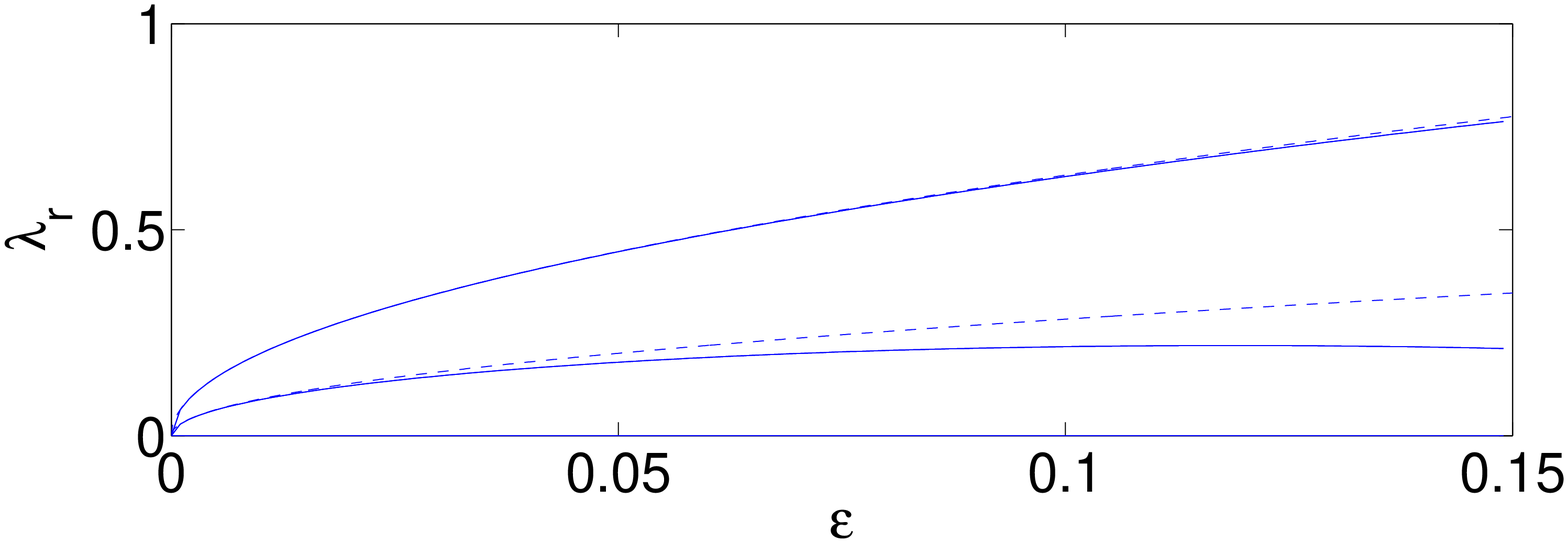}}
\caption{The top panel shows the spatial profile of the mode ++,++
in both components
and the corresponding spectral plane of the linear stability problem
is shown in the middle panel
for $\epsilon = 0.05$. The bottom subplot shows the continuation of the branch from
$\epsilon = 0$ to $\epsilon = 0.15$ and the real positive eigenvalue
numerically obtained (solid line) and theoretically predicted
(dashed line). The theory yields
$\lambda^2=4\epsilon,\  \frac{4}{5}\epsilon$.}
\label{fig++,++}
\end{figure}

\begin{figure}[htp]
\scalebox{0.375}{\includegraphics{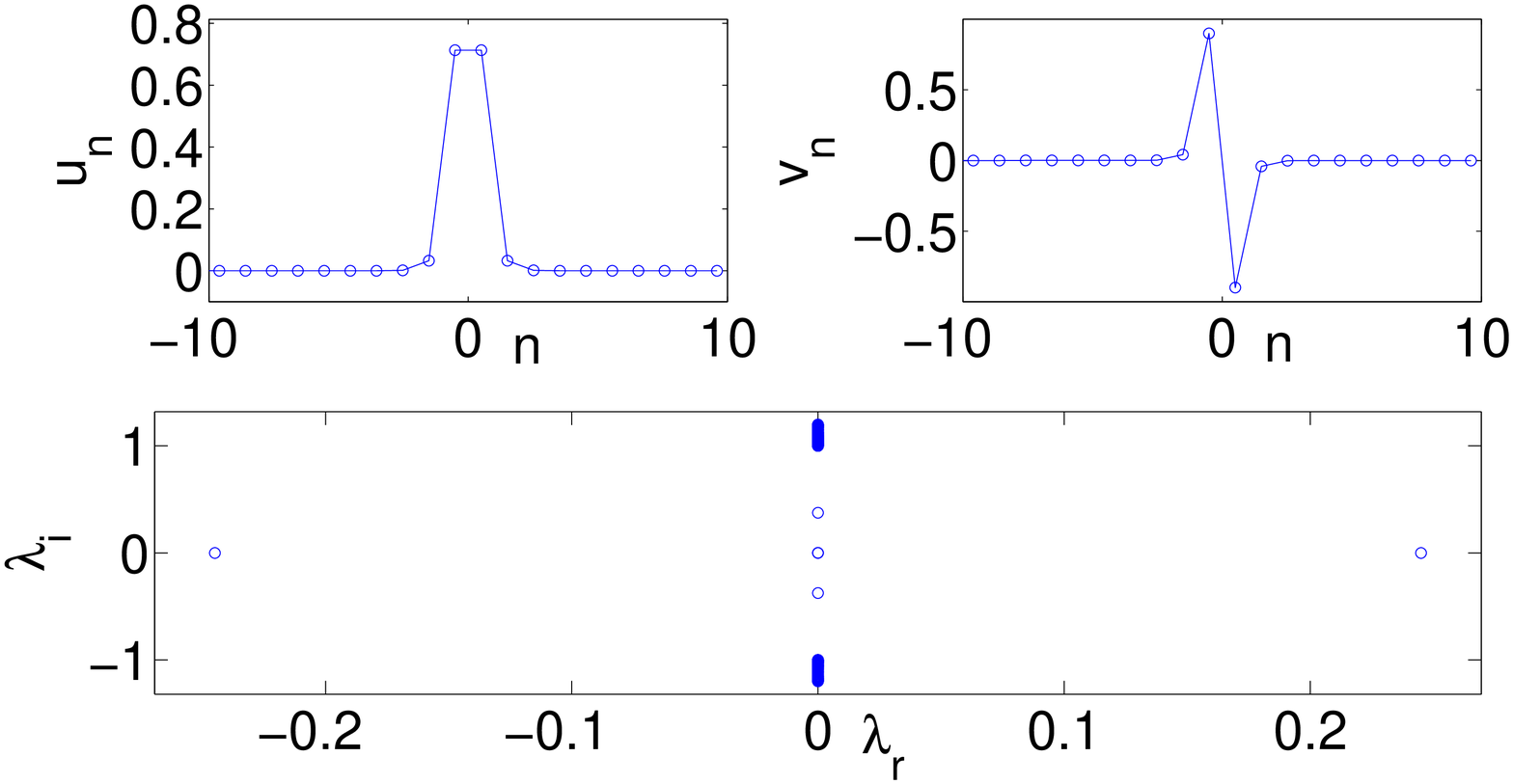}}
\scalebox{0.375}{\includegraphics{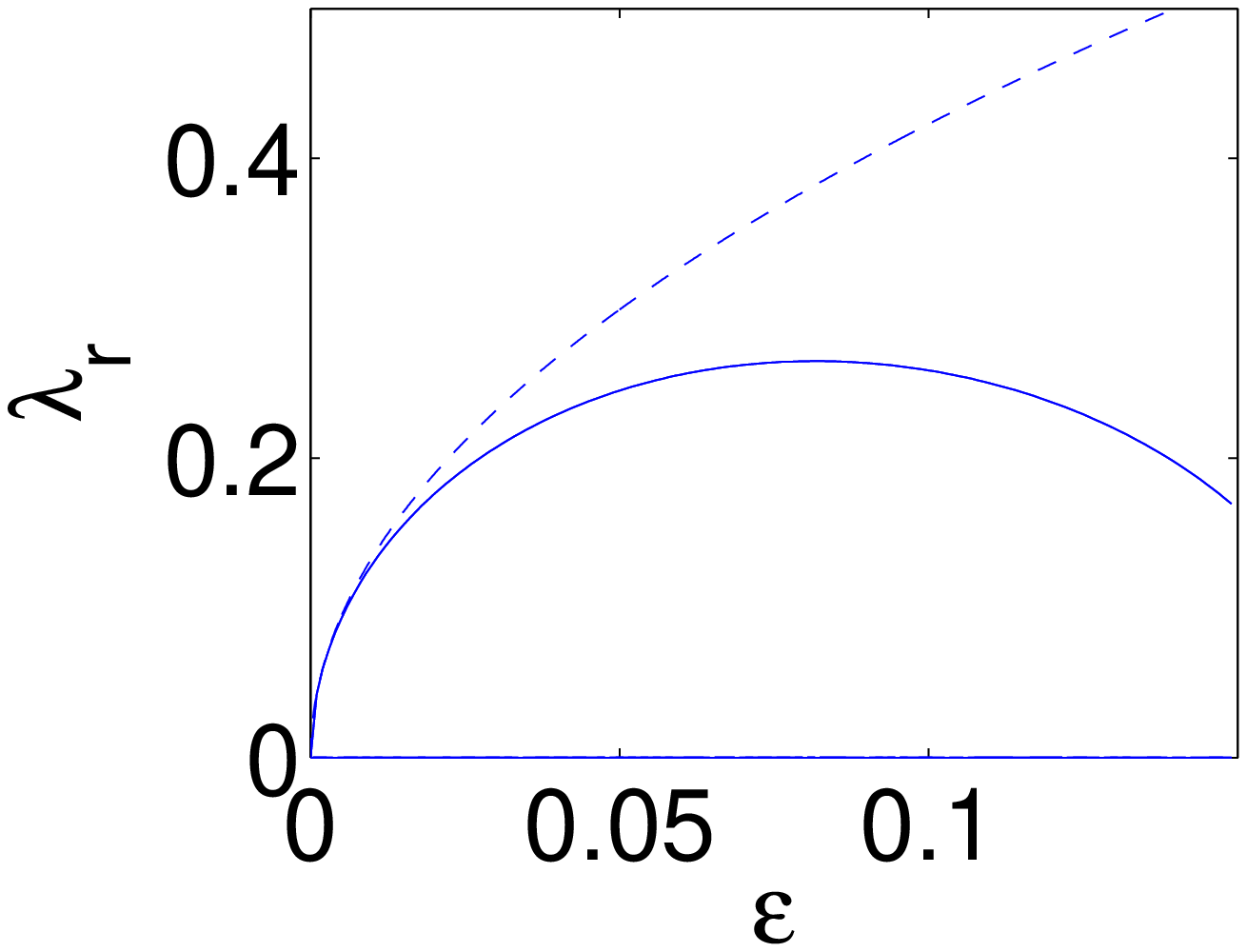}}
\scalebox{0.375}{\includegraphics{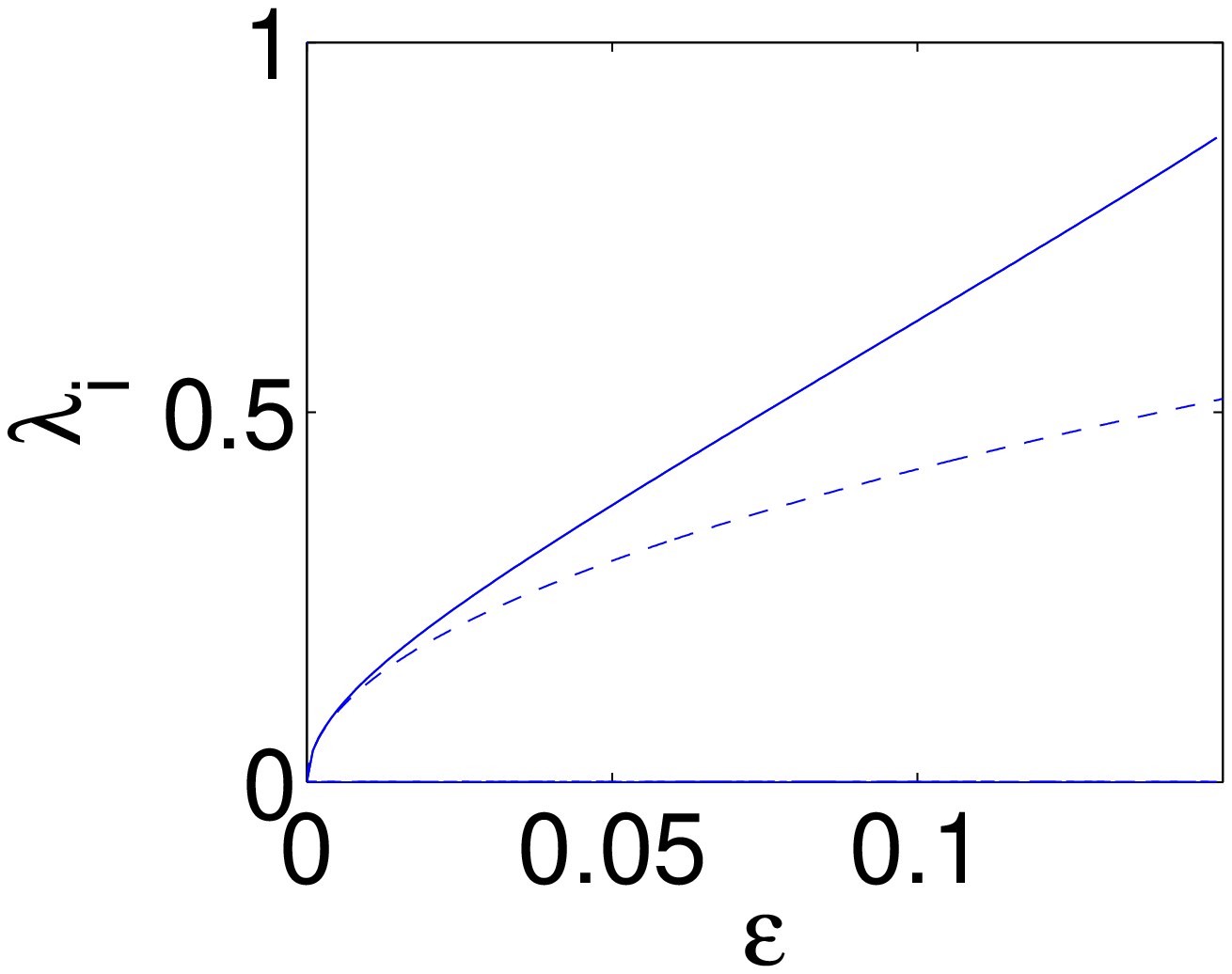}}
\caption{The top panel shows the mode ++,+- and the middle panel
its spectral plane for $\epsilon = 0.05$.
The bottom subplot shows the real and imaginary parts of the
eigenvalues
(again numerics in solid vs. theory in dashed). The theoretical
prediction gives
$\lambda^2=\pm\frac{4}{\sqrt{5}}\epsilon$.}
\label{fig++,+-}
\end{figure}

\begin{figure}[htp]
\scalebox{0.375}{\includegraphics{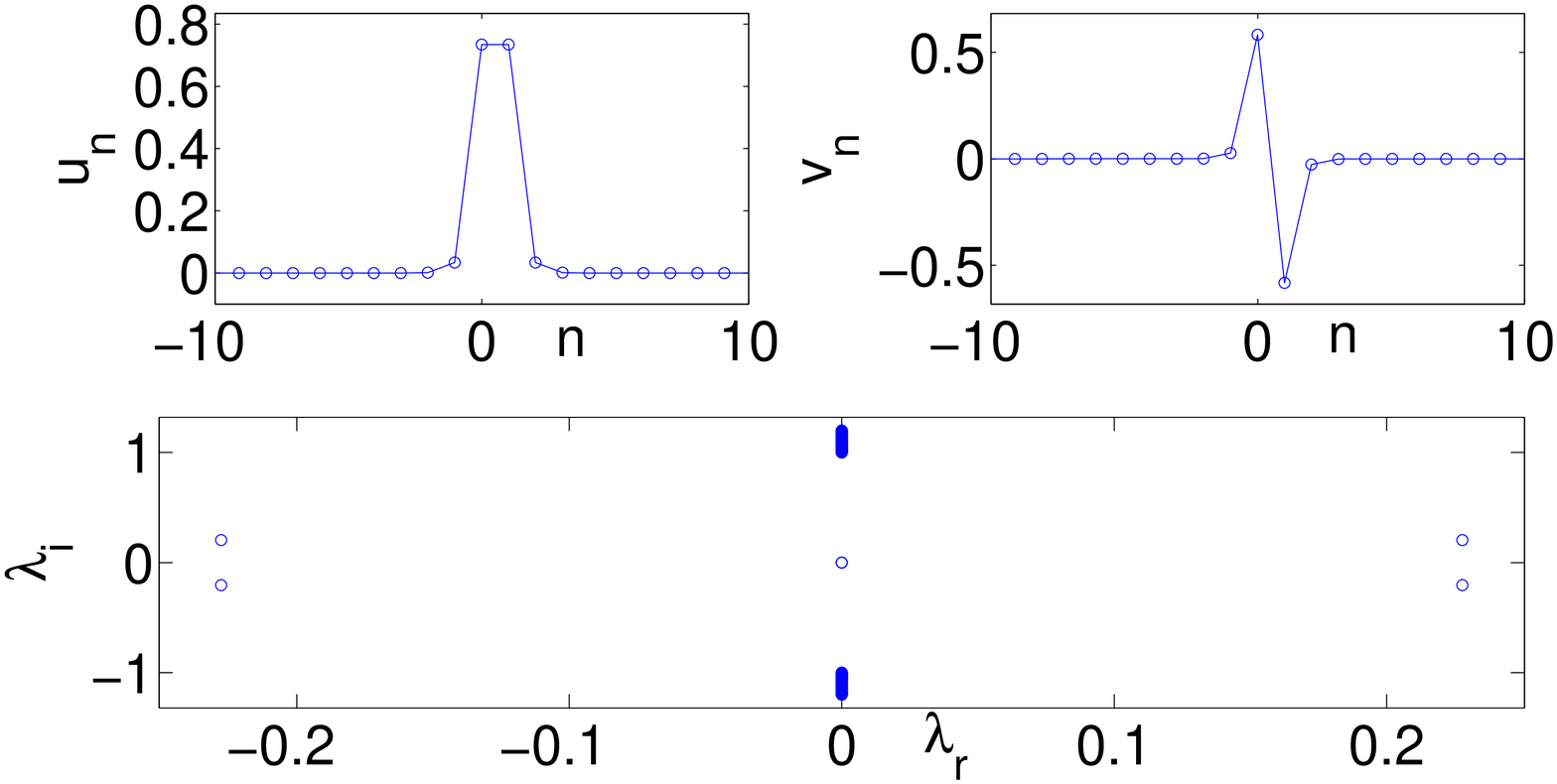}}
\scalebox{0.375}{\includegraphics{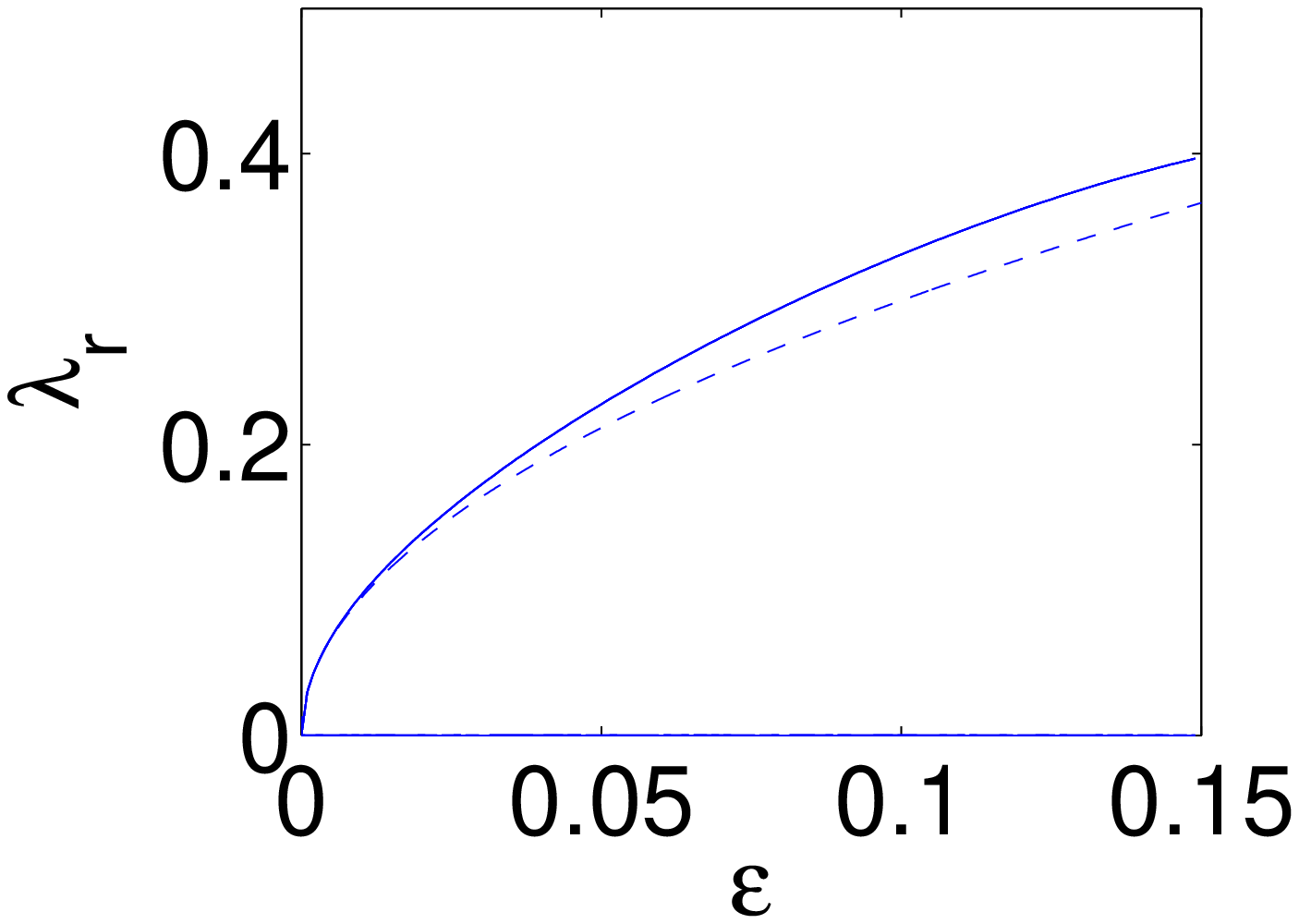}}
\scalebox{0.375}{\includegraphics{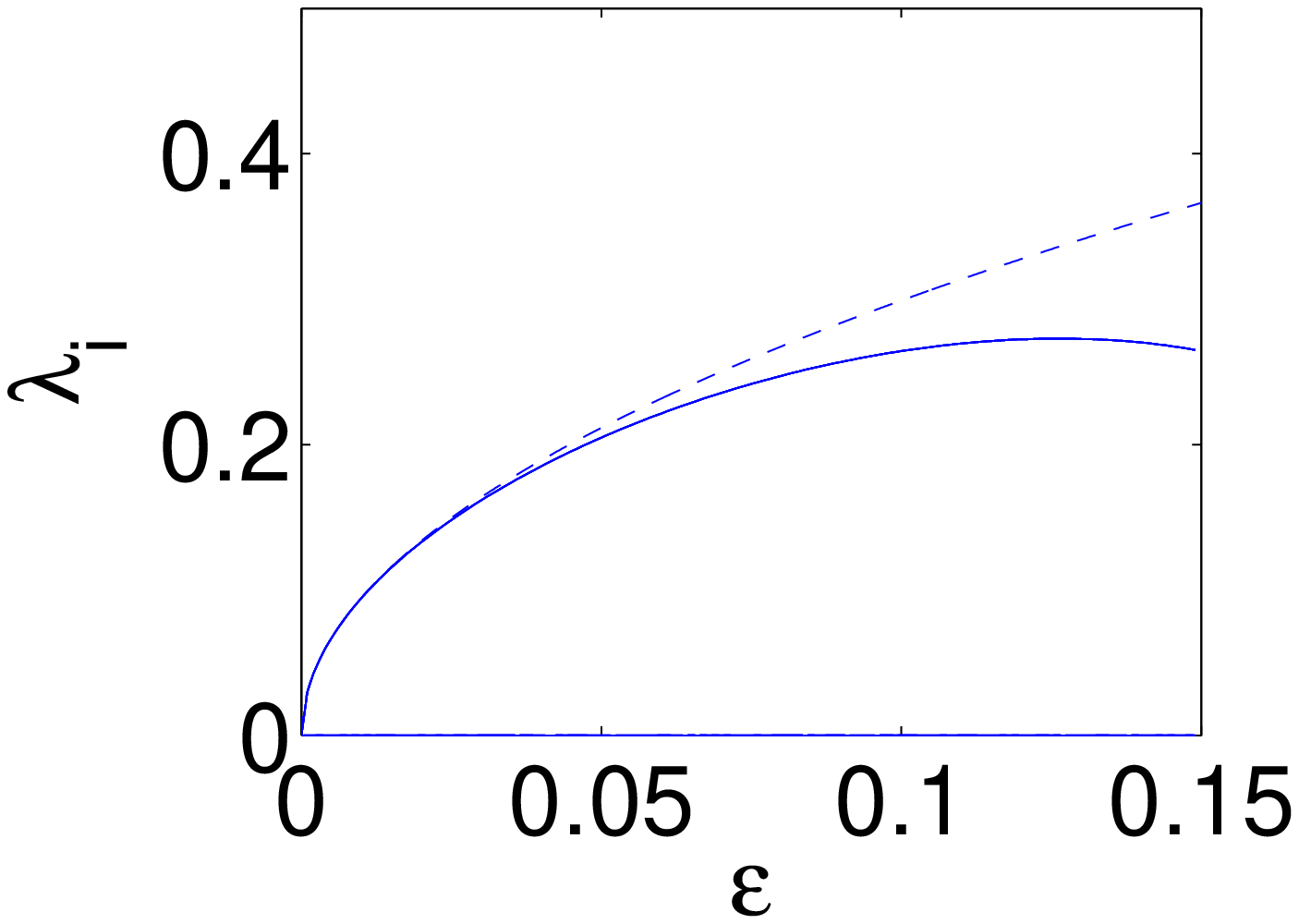}}
\caption{The top
panels show the mode ++,+- but with $g_{11}=g_{22}=1,\ g_{12}=1.5$
and the middle panel its spectral plane for $\epsilon = 0.05$.
The bottom subplot shows the real and imaginary part of the quartet, for
which the theoretical prediction is
$\lambda^2=\pm\frac{4i}{\sqrt{5}}\epsilon$.}
\label{fig++,+-_g12}
\end{figure}

\begin{figure}[htp]
\scalebox{0.375}{\includegraphics{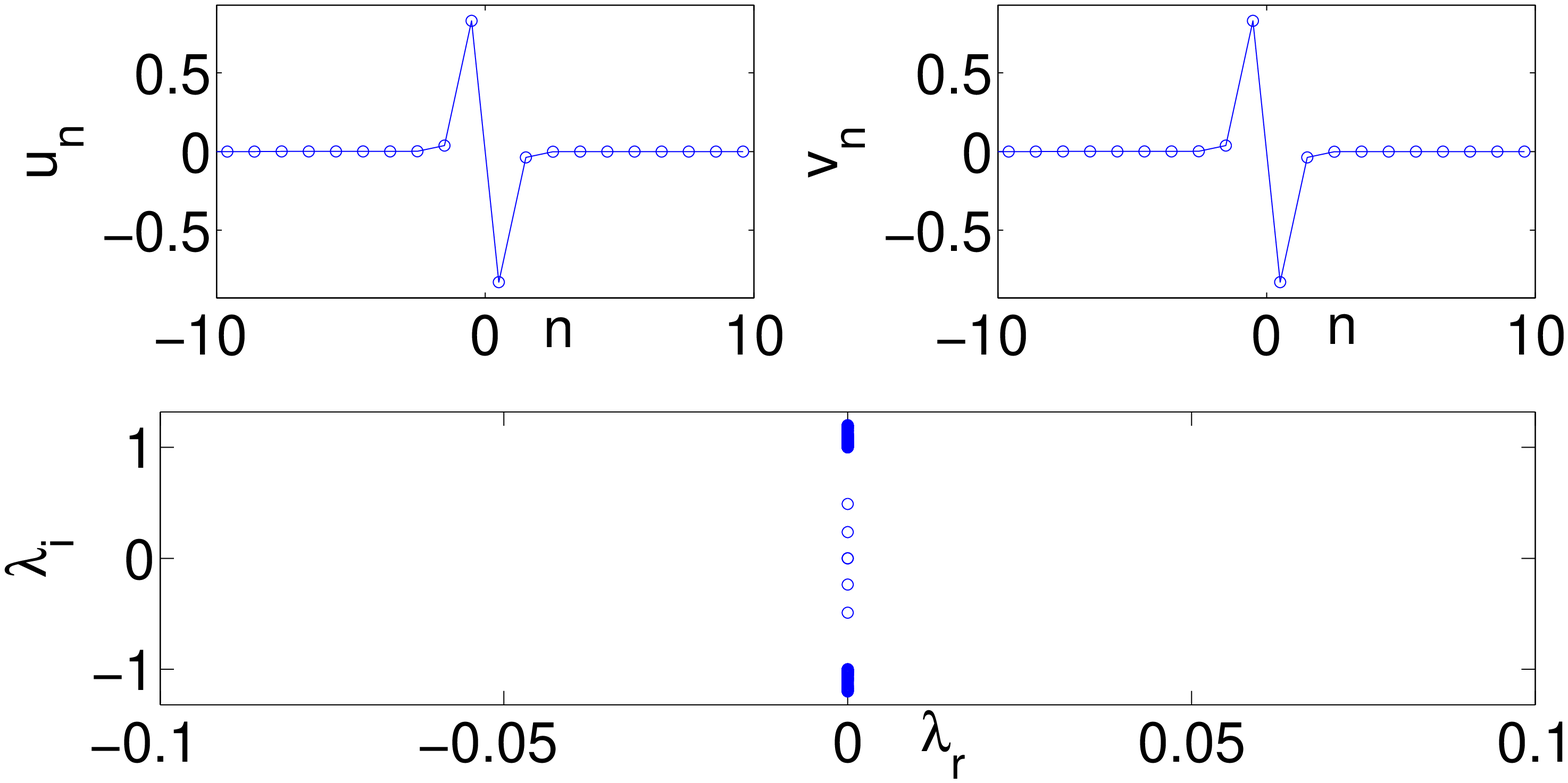}}
\scalebox{0.375}{\includegraphics{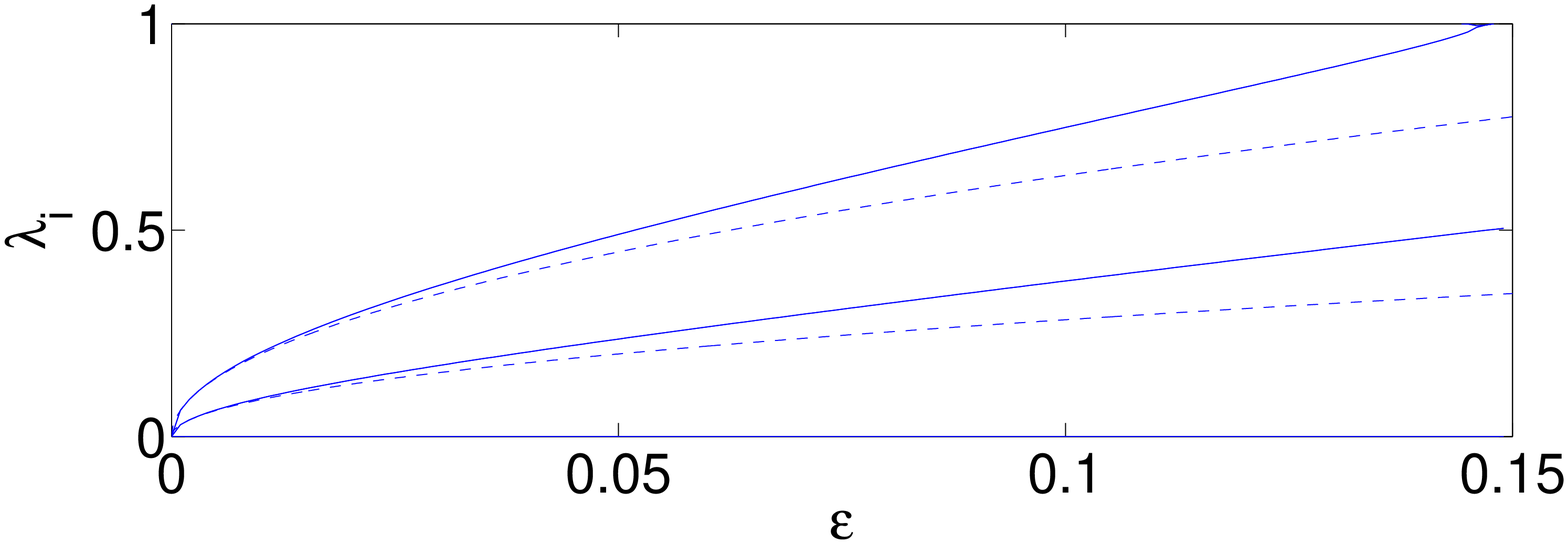}}
\caption{The top panel shows the mode +-,+- and the spectral plane for $\epsilon = 0.05$.
The bottom subplot shows the imaginary part of the eigenvalues, for which
the theoretical
analysis yields $\lambda^2=-4\epsilon,\  -\frac{4}{5}\epsilon$.}
\label{fig+-,+-}
\end{figure}

\begin{figure}[htp]
\scalebox{0.375}{\includegraphics{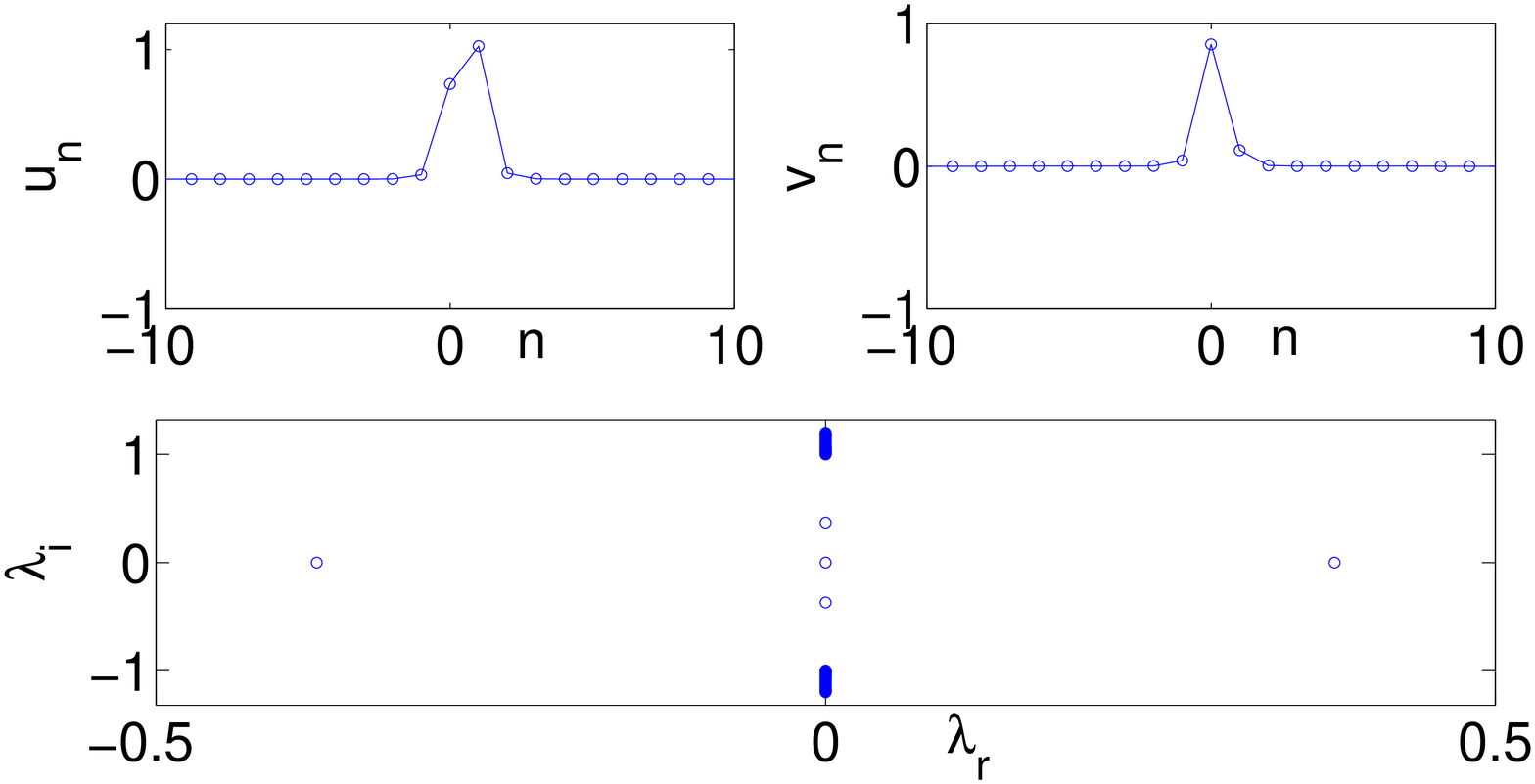}}
\scalebox{0.375}{\includegraphics{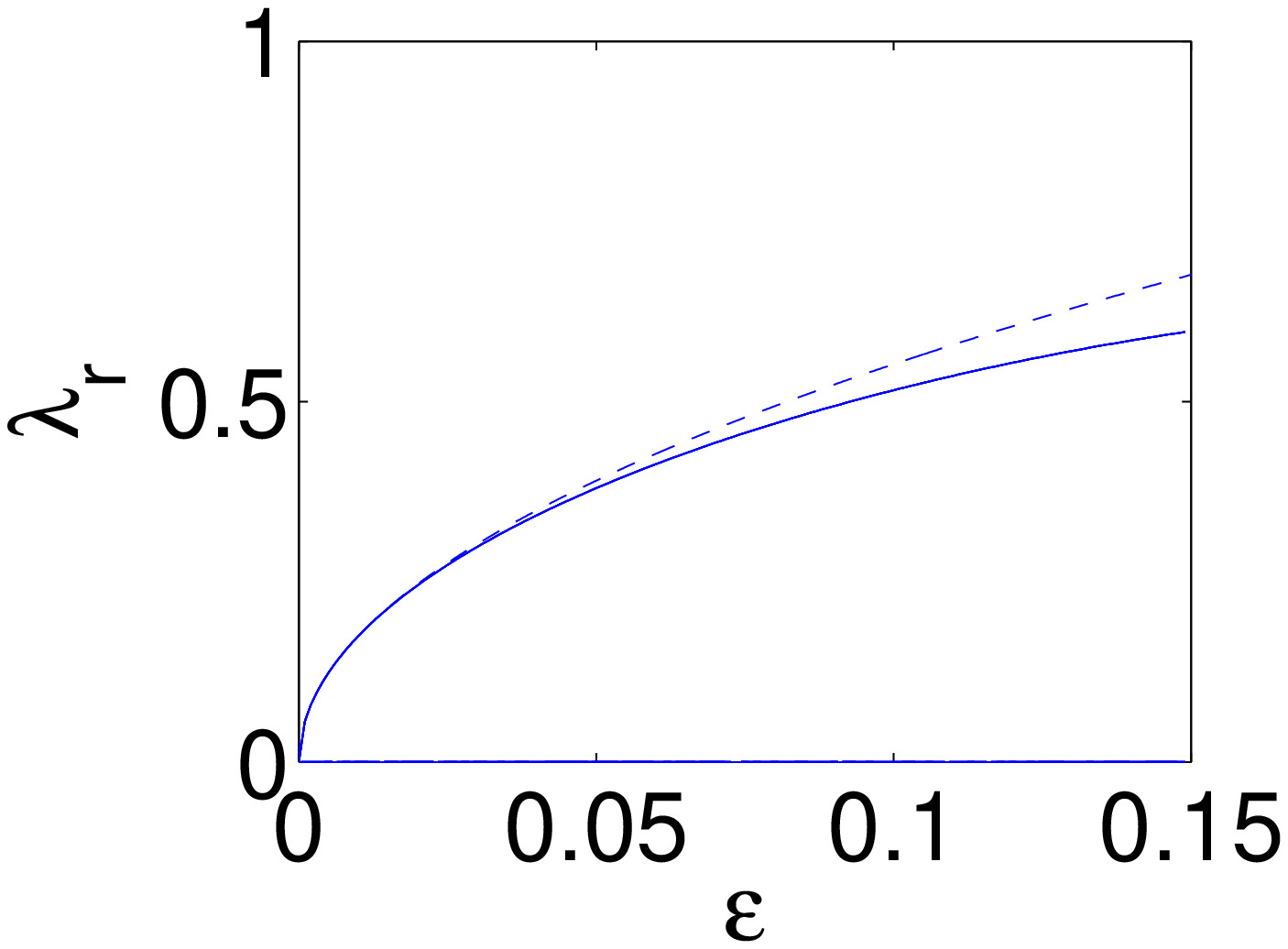}}
\scalebox{0.375}{\includegraphics{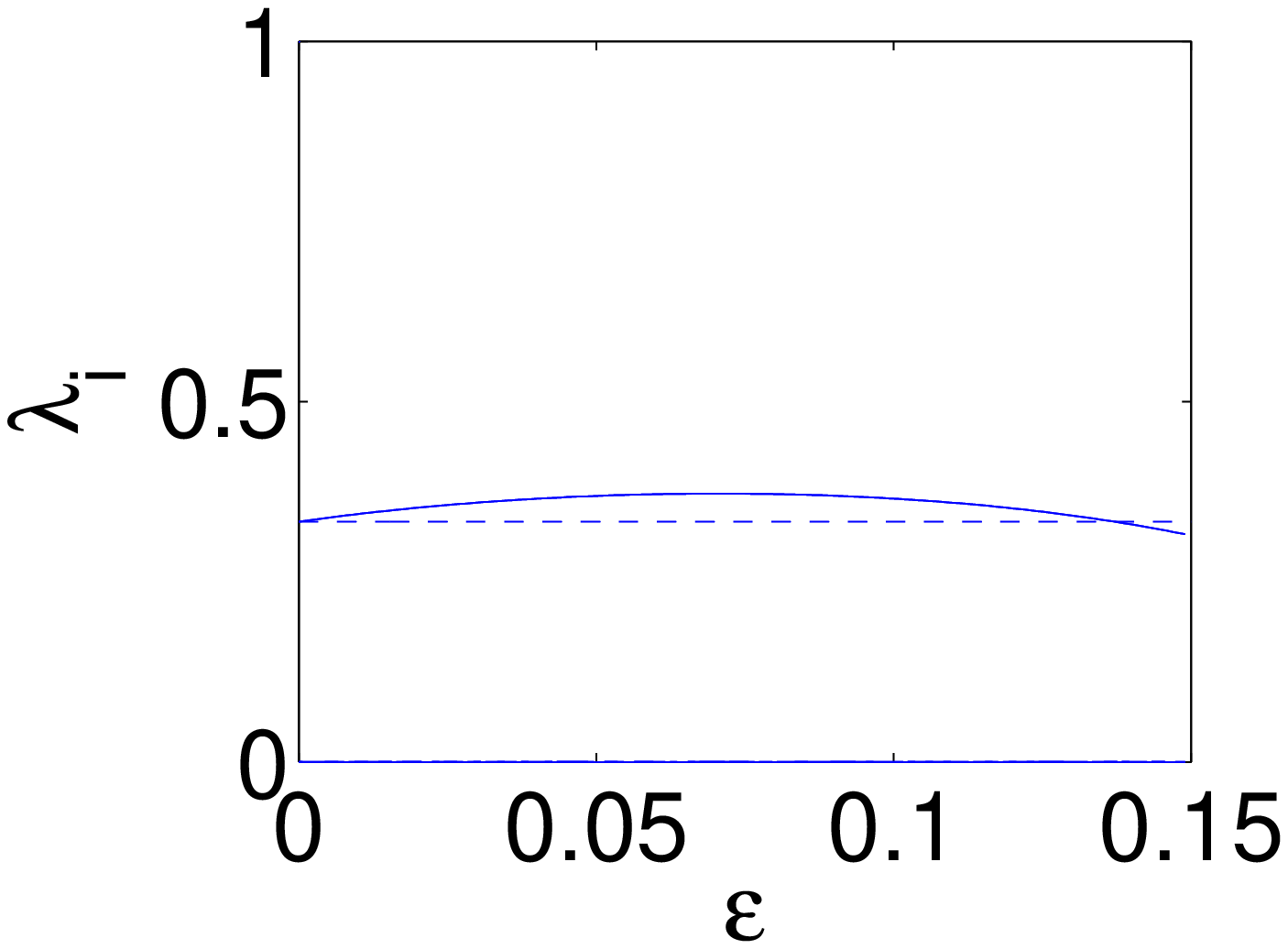}}
\caption{The top panel shows the mode ++,+0 and the spectral plane for $\epsilon = 0.05$.
The bottom subplot shows the real and imaginary parts of the eigenvalues given
(theoretically) by
$\lambda=\frac{1}{3}i,\  1.74543\sqrt{\epsilon}$.}
\label{fig++,+0}
\end{figure}

\begin{figure}[htp]
\scalebox{0.375}{\includegraphics{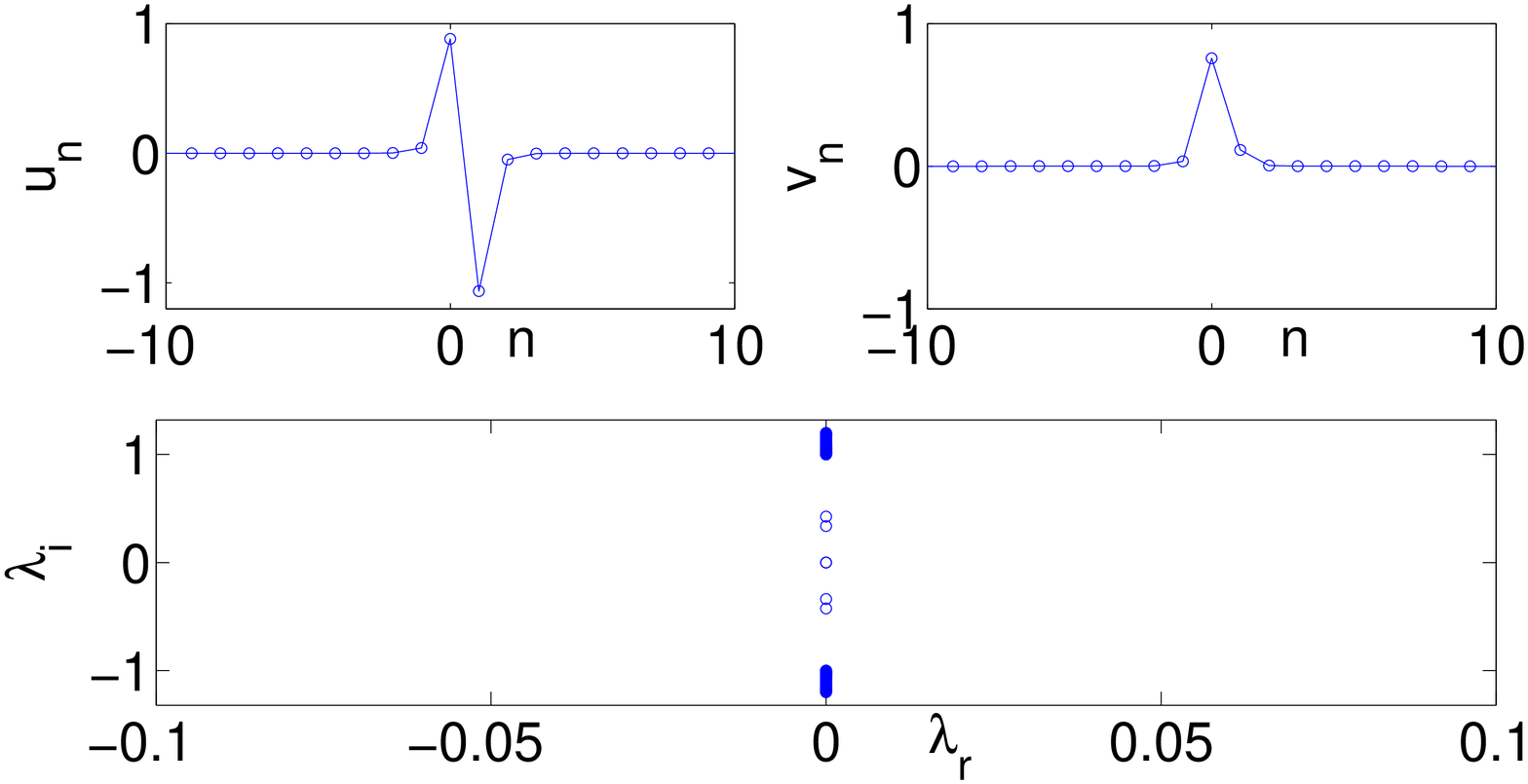}}
\scalebox{0.375}{\includegraphics{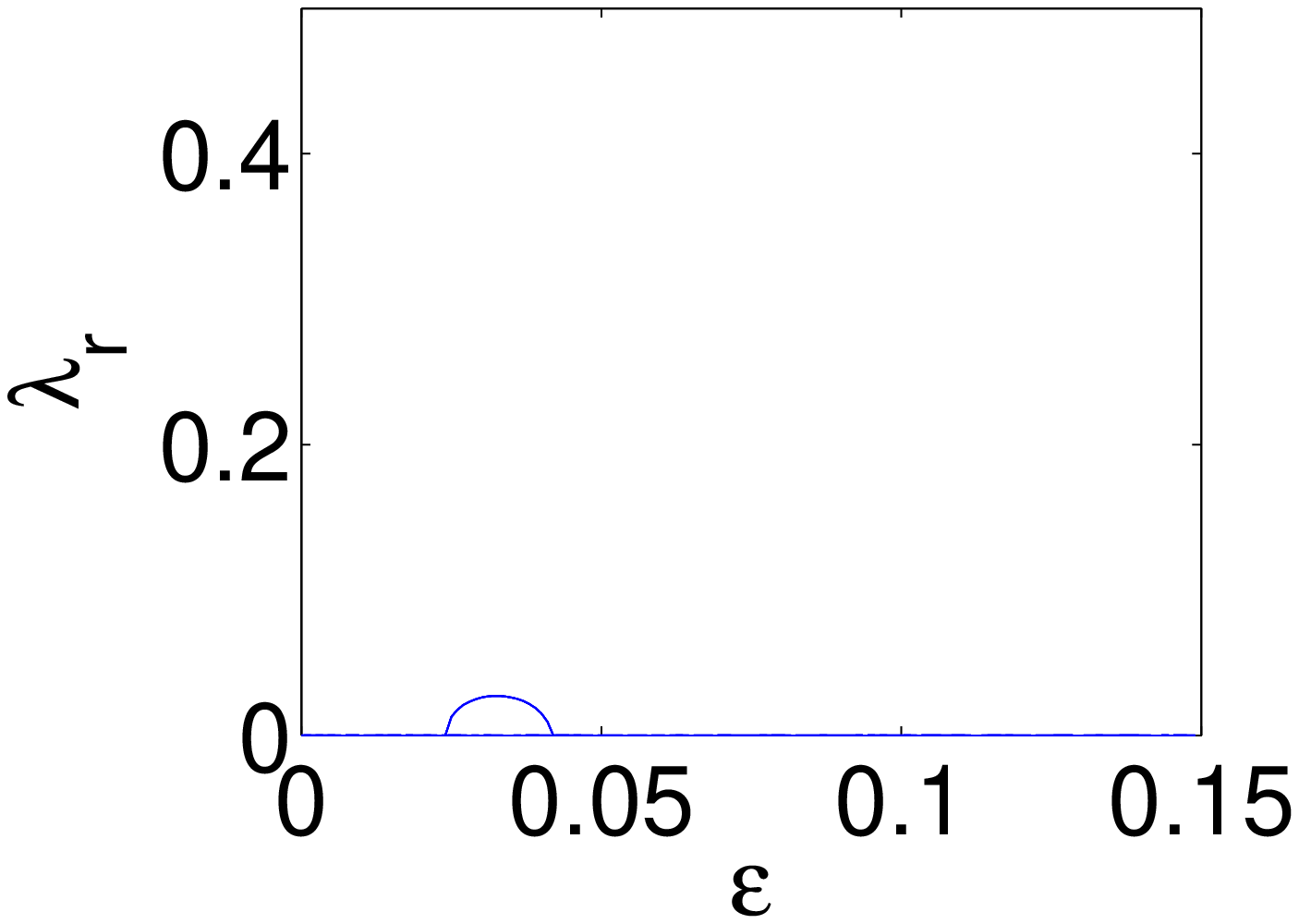}}
\scalebox{0.375}{\includegraphics{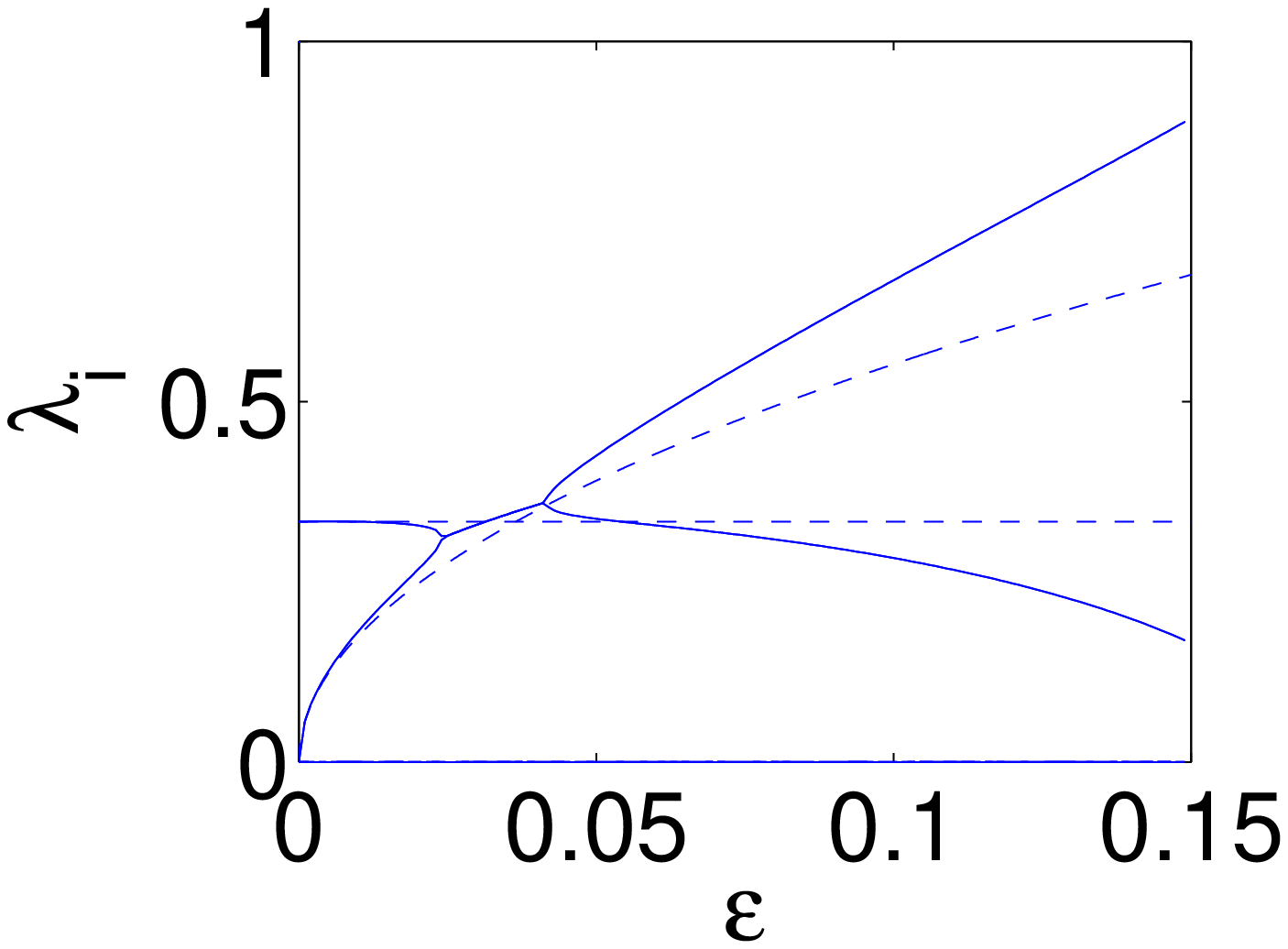}}
\caption{The top panel shows the mode +-,+0 and the spectral plane for $\epsilon = 0.05$.
The bottom subplot shows the two pairs of purely imaginary eigenvalues
theoretically predicted as
$\lambda=1.74543\sqrt{\epsilon}i,\  \frac{1}{3}i,$. They collide from
$\epsilon=0.025$
to $\epsilon=0.041$, i.e., the oscillatory instability is only
observed for intermediate values of $\epsilon$ and disappears
for larger values of $\epsilon$.}
\label{fig+-,+0}
\end{figure}

\begin{figure}[htp]
\centering
\scalebox{0.375}{\includegraphics{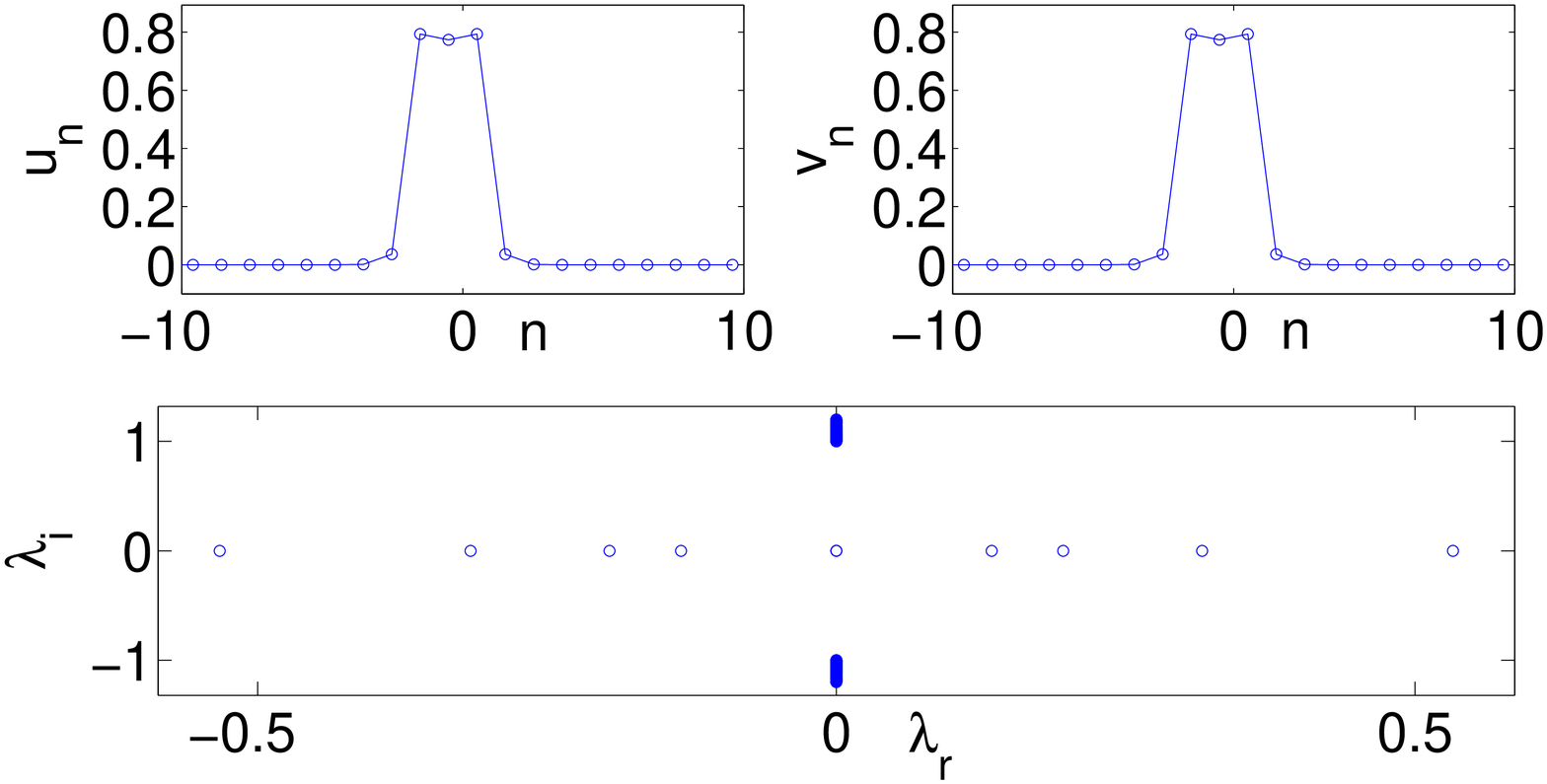}}
\scalebox{0.375}{\includegraphics{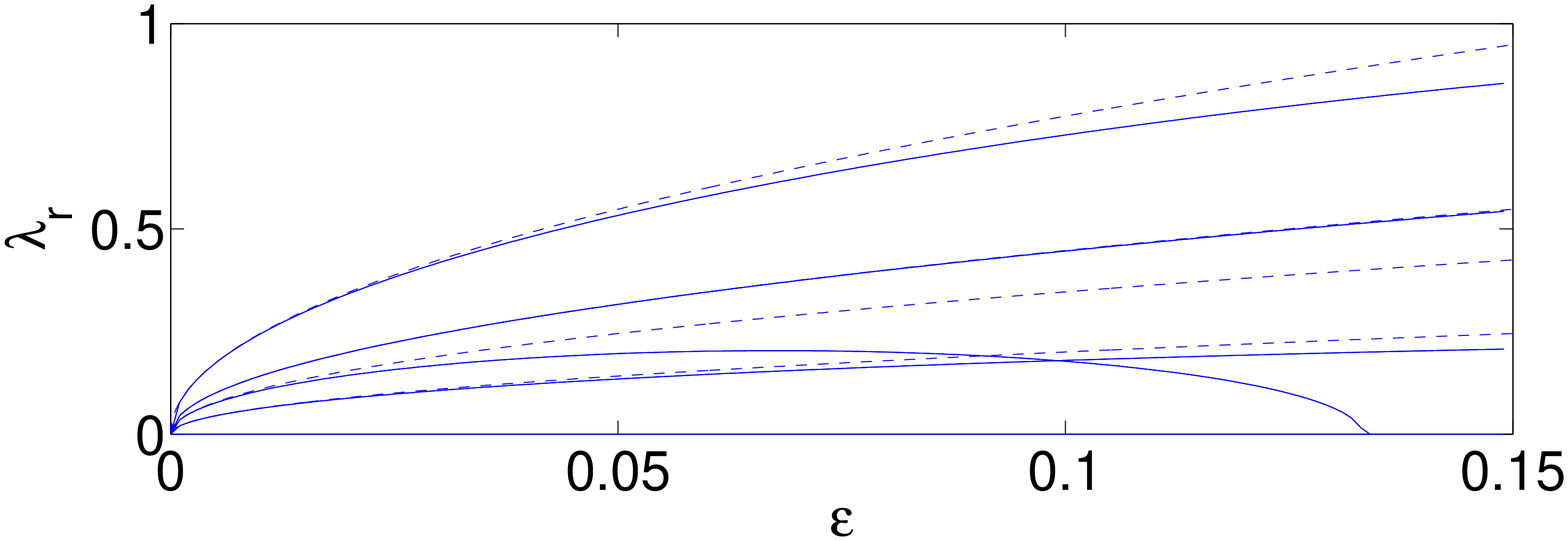}}
\caption{The mode +++,+++ (top panel) with four pairs of real
  eigenvalues
(see the spectral plane of the middle panel) is shown in the
figure, captured when $\epsilon = 0.05$. The analytical prediction for these pairs
$\lambda^2=6\epsilon,\  2\epsilon,\  \frac{6}{5}\epsilon,\
\frac{2}{5}\epsilon$ is shown by dashed line in the bottom panel
and compared to the numerical results (solid lines).}
\label{fig+++,+++}
\end{figure}

\begin{figure}[htp]
\centering
\scalebox{0.375}{\includegraphics{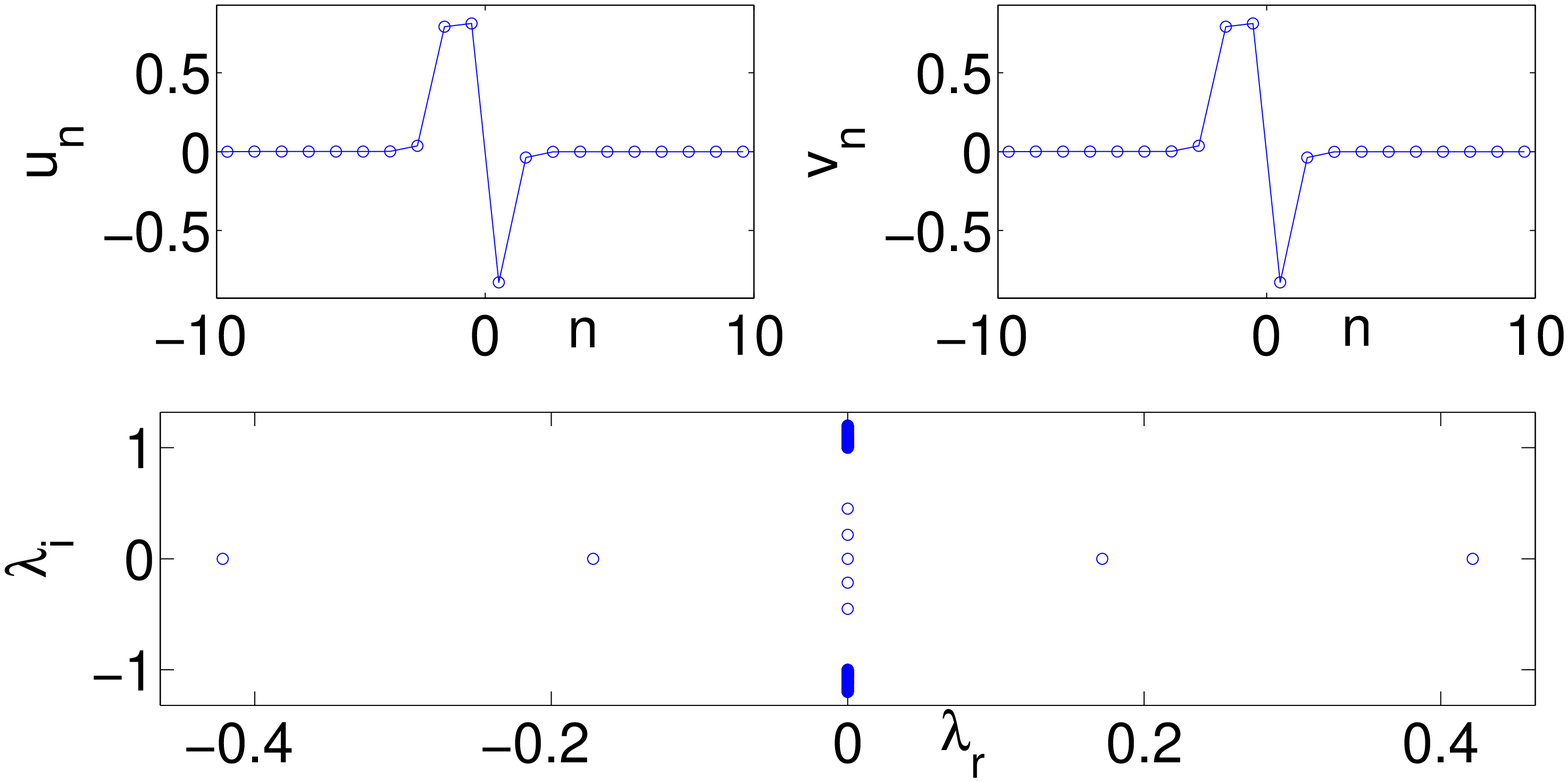}}
\scalebox{0.375}{\includegraphics{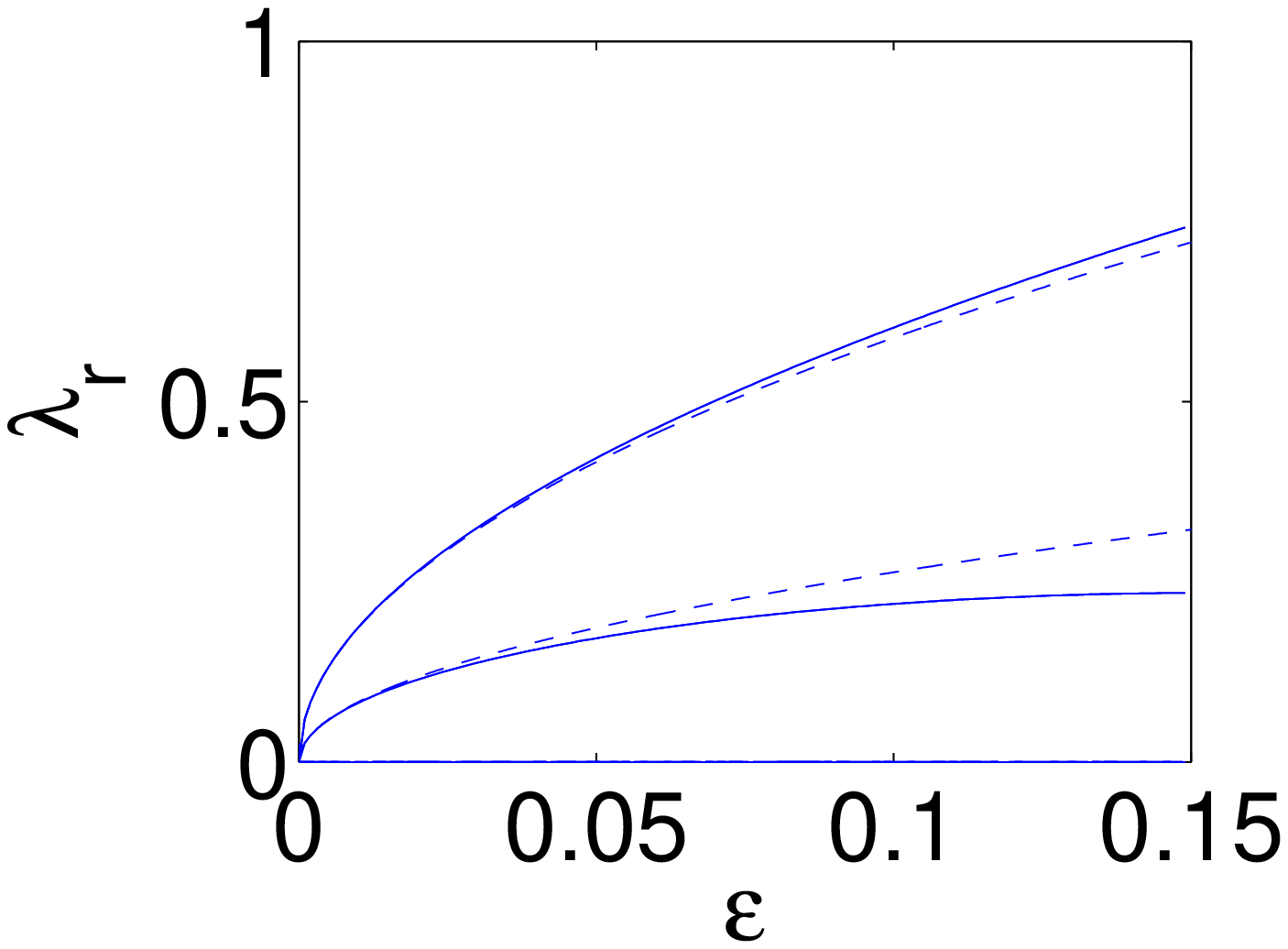}}
\scalebox{0.375}{\includegraphics{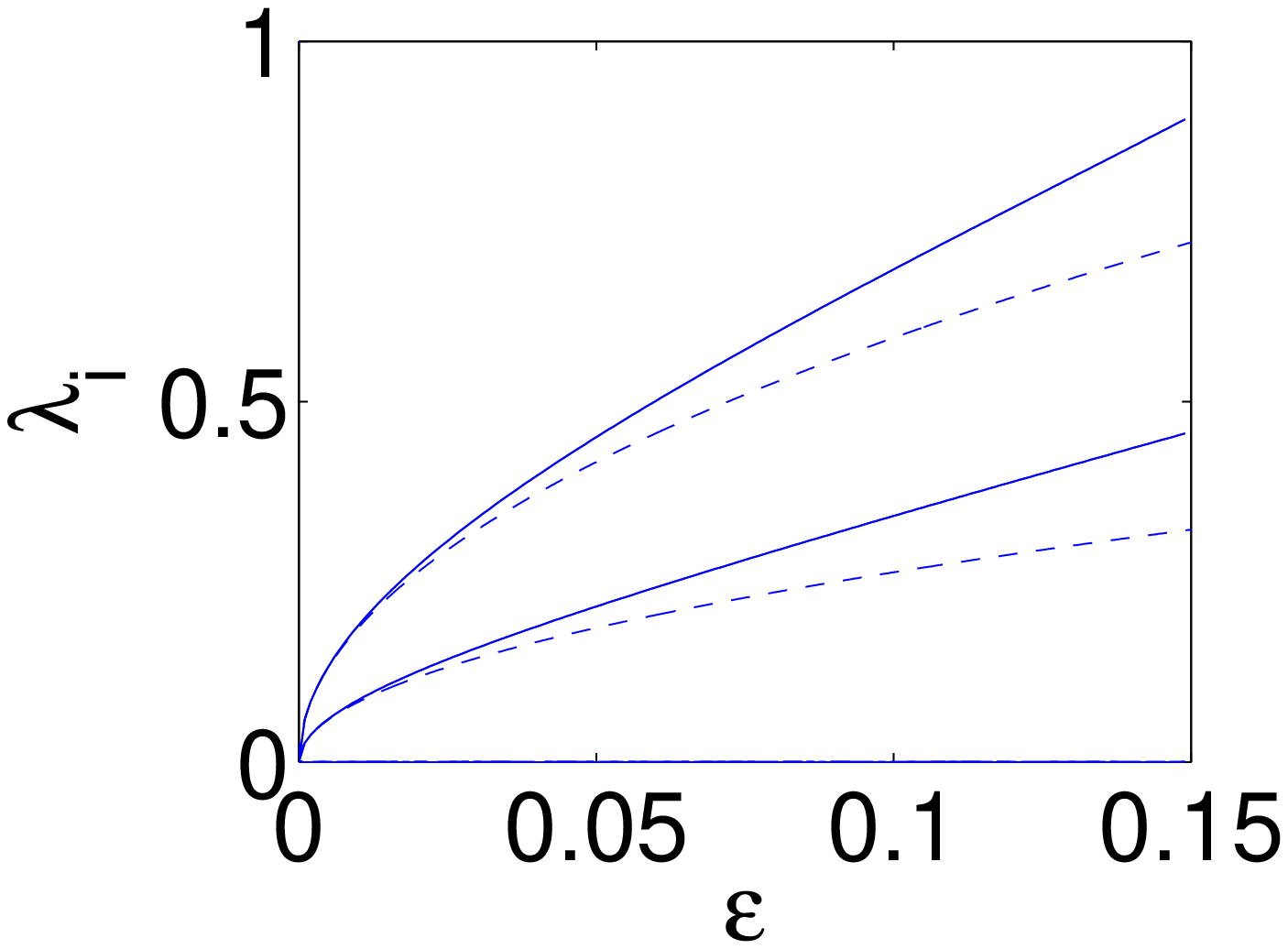}}
\caption{Similarly to Fig~\ref{fig+++,+++} but for the mode ++-,++- with
two pairs of real and two pairs of purely imaginary eigenvalues given by
$\lambda^2=\pm2\sqrt{3}\epsilon,\  \pm\frac{2\sqrt{3}}{5}\epsilon$.
The top three panels are captured when $\epsilon = 0.05$.}
\label{fig++-,++-}
\end{figure}

\begin{figure}[htp]
\centering
\scalebox{0.375}{\includegraphics{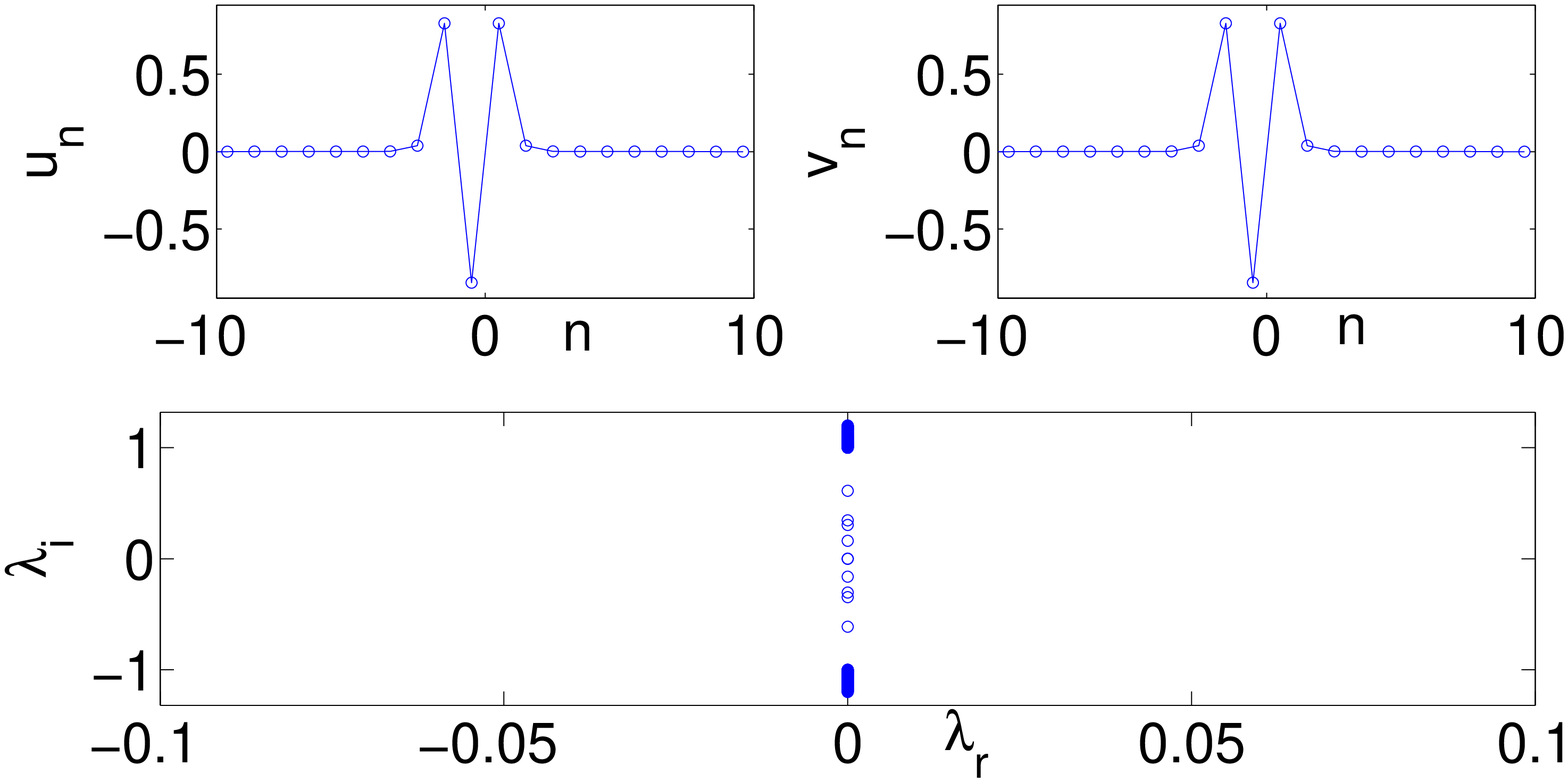}}
\scalebox{0.375}{\includegraphics{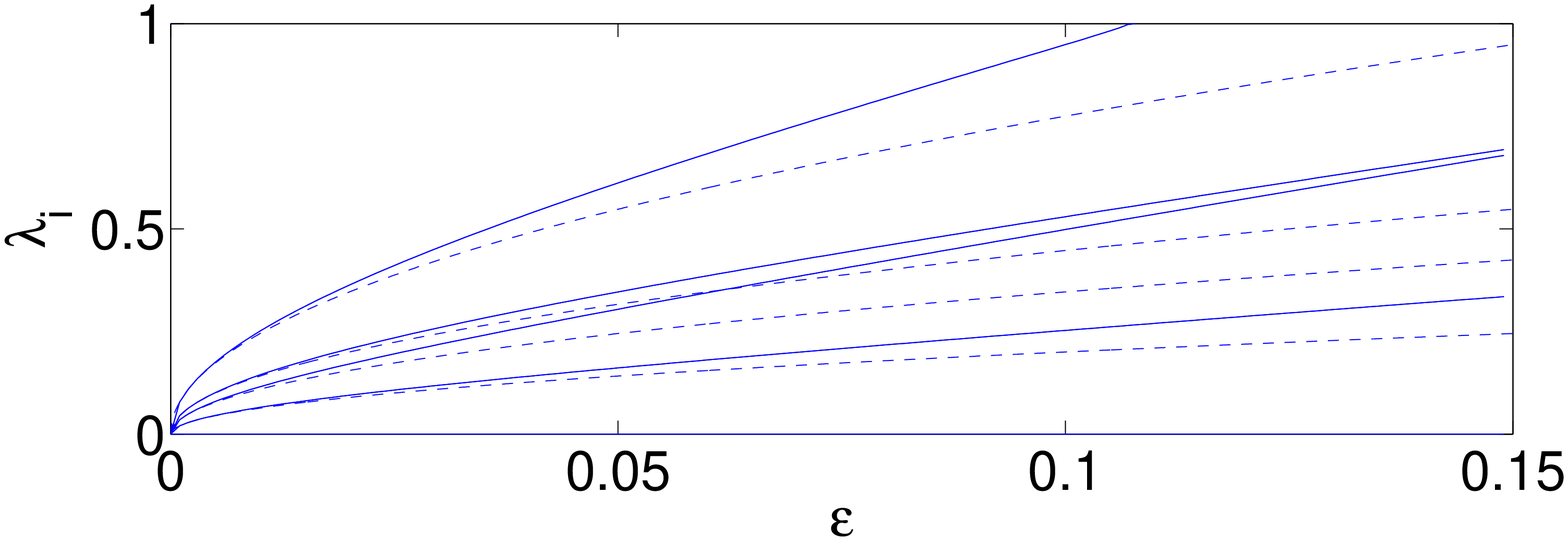}}
\caption{Similarly to Fig~\ref{fig+++,+++} but for the mode +-+,+-+
with four pairs of purely imaginary eigenvalues theoretically given by
$\lambda^2=-6\epsilon,\  -2\epsilon,\  -\frac{6}{5}\epsilon,\  -\frac{2}{5}\epsilon$.
The top three panels are captured when $\epsilon = 0.05$.}
\label{fig+-+,+-+}
\end{figure}

\begin{figure}[htp]
\scalebox{0.375}{\includegraphics{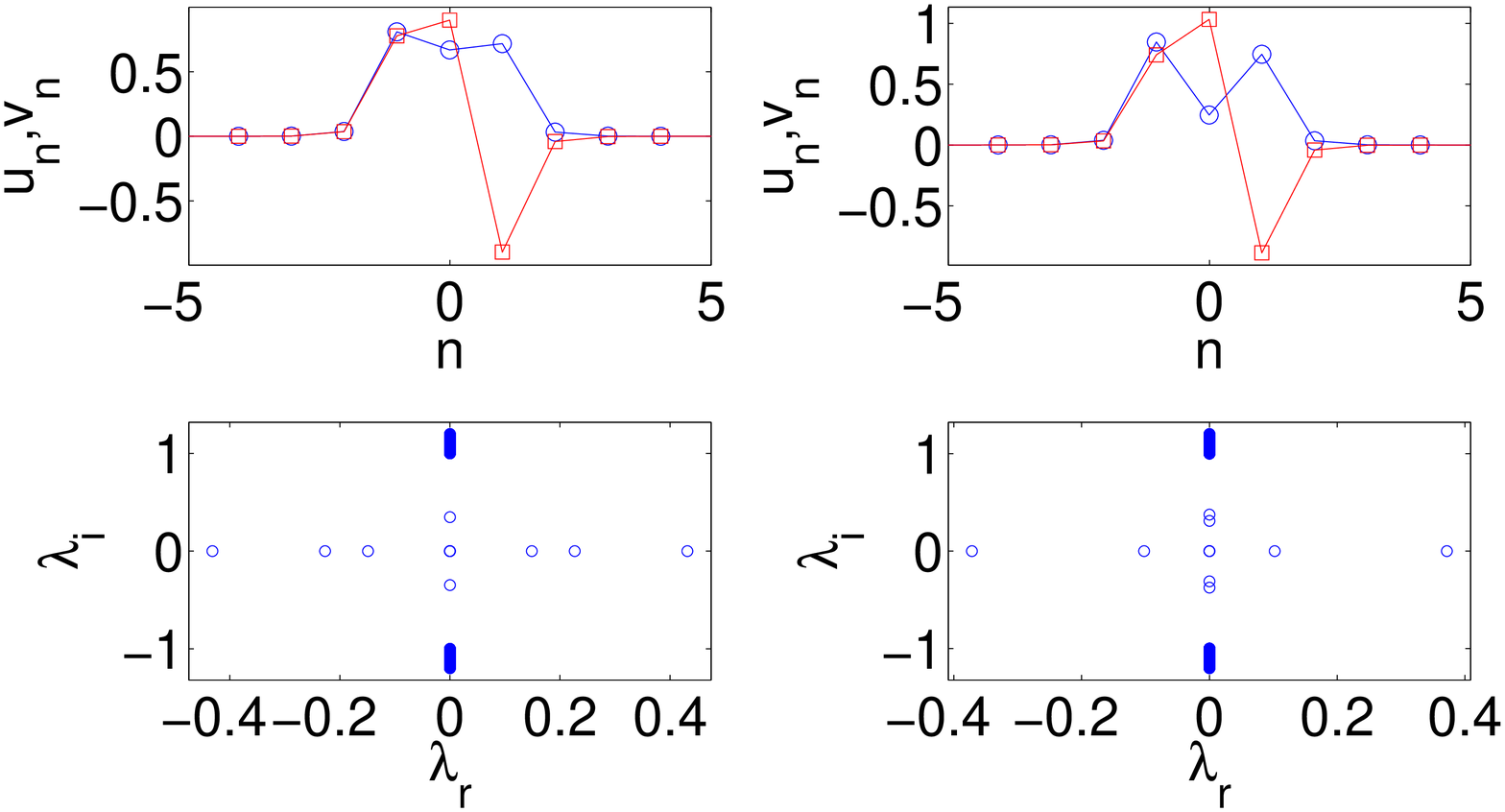}}
\scalebox{0.375}{\includegraphics{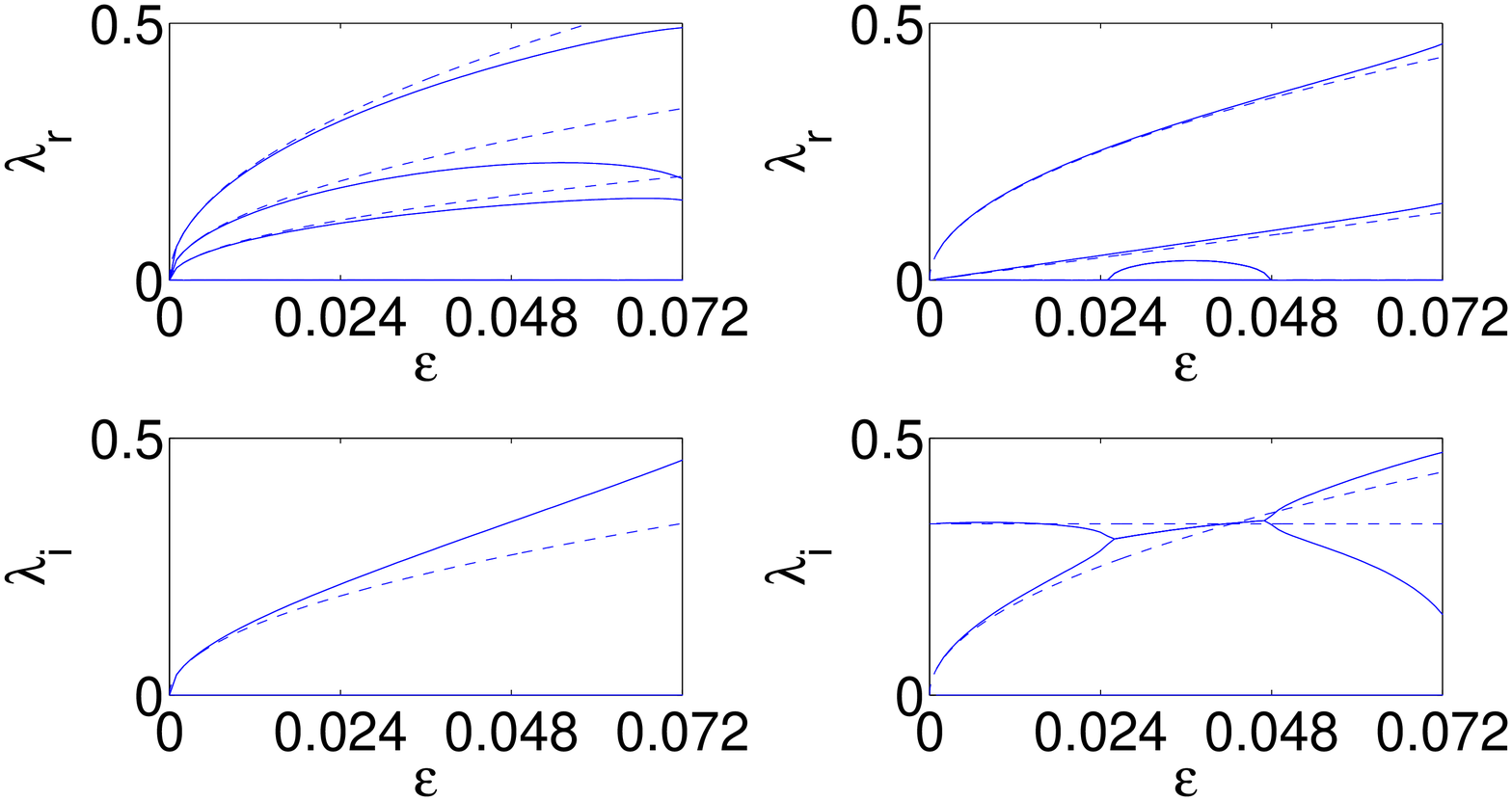}}
\caption{The left column shows the mode +++,++-, which terminates
 at $\epsilon=0.072$ by colliding with the mode +0+,++- shown in the
 right
column. The top two
rows show the profile and the linear
stability of each mode when $\epsilon=0.05$. The bottom two rows
show the eigenvalues of each mode. The mode +++,++- has three pairs of real and
one pair of purely imaginary eigenvalues given by $\lambda^2=\frac{2}{5} \left(6\pm\sqrt{21}\right)\epsilon,
\ \pm2 \sqrt{\frac{3}{5}}\epsilon$, while the mode +0+,++-
has two pairs of real and
two pair of purely imaginary eigenvalues given by $\lambda=\frac{1}{3}i,\ 1.61812\sqrt{\epsilon}i,\ 1.61812\sqrt{\epsilon},
\ 1.82599\epsilon$. The first two pairs of eigenvalues collide and
form a complex eigenvalue quartet from $\epsilon=0.025$ to $0.048$.}
\label{fig+++,++-}
\end{figure}

\begin{figure}[htp]
\scalebox{0.375}{\includegraphics{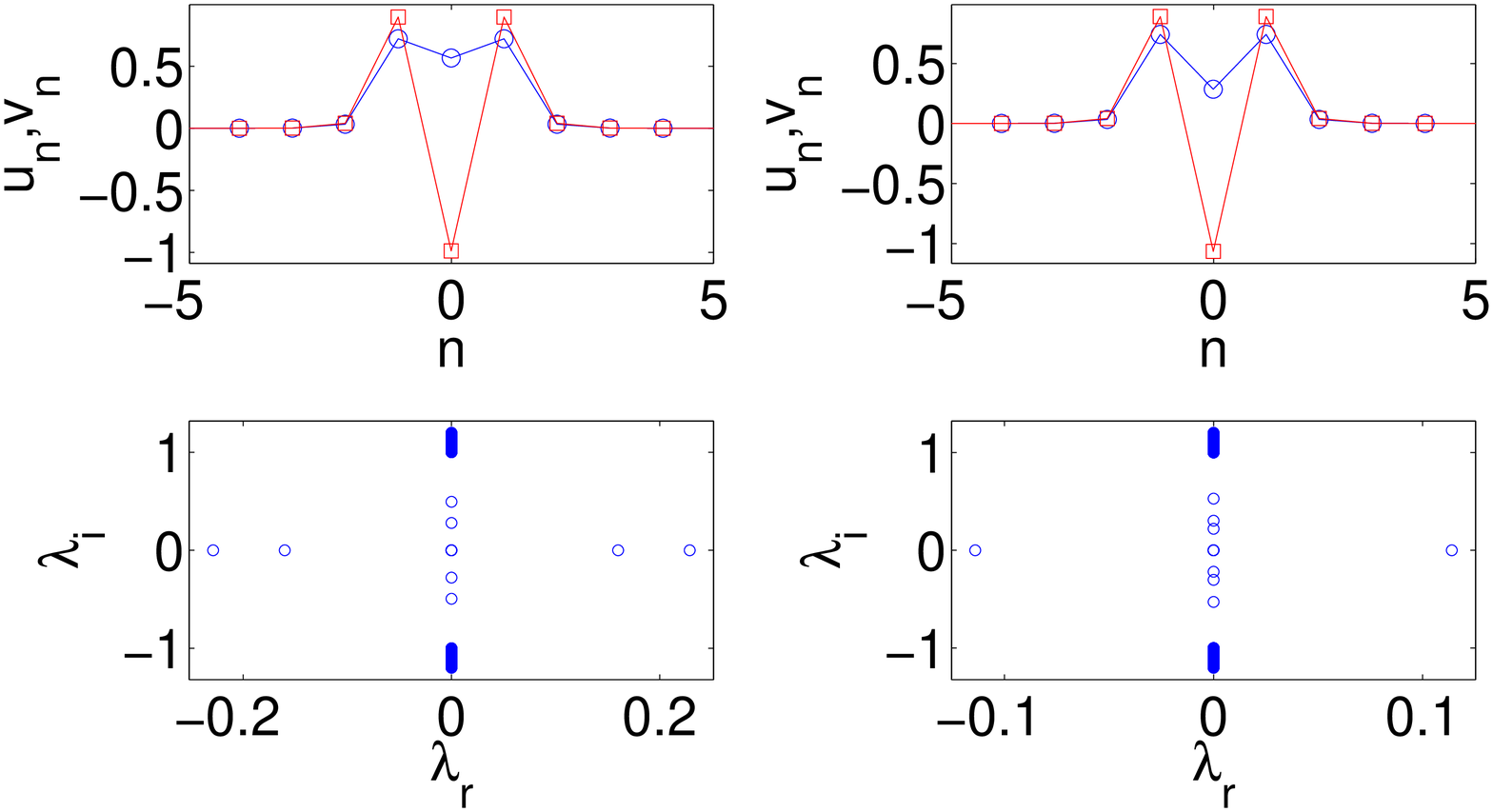}}
\scalebox{0.375}{\includegraphics{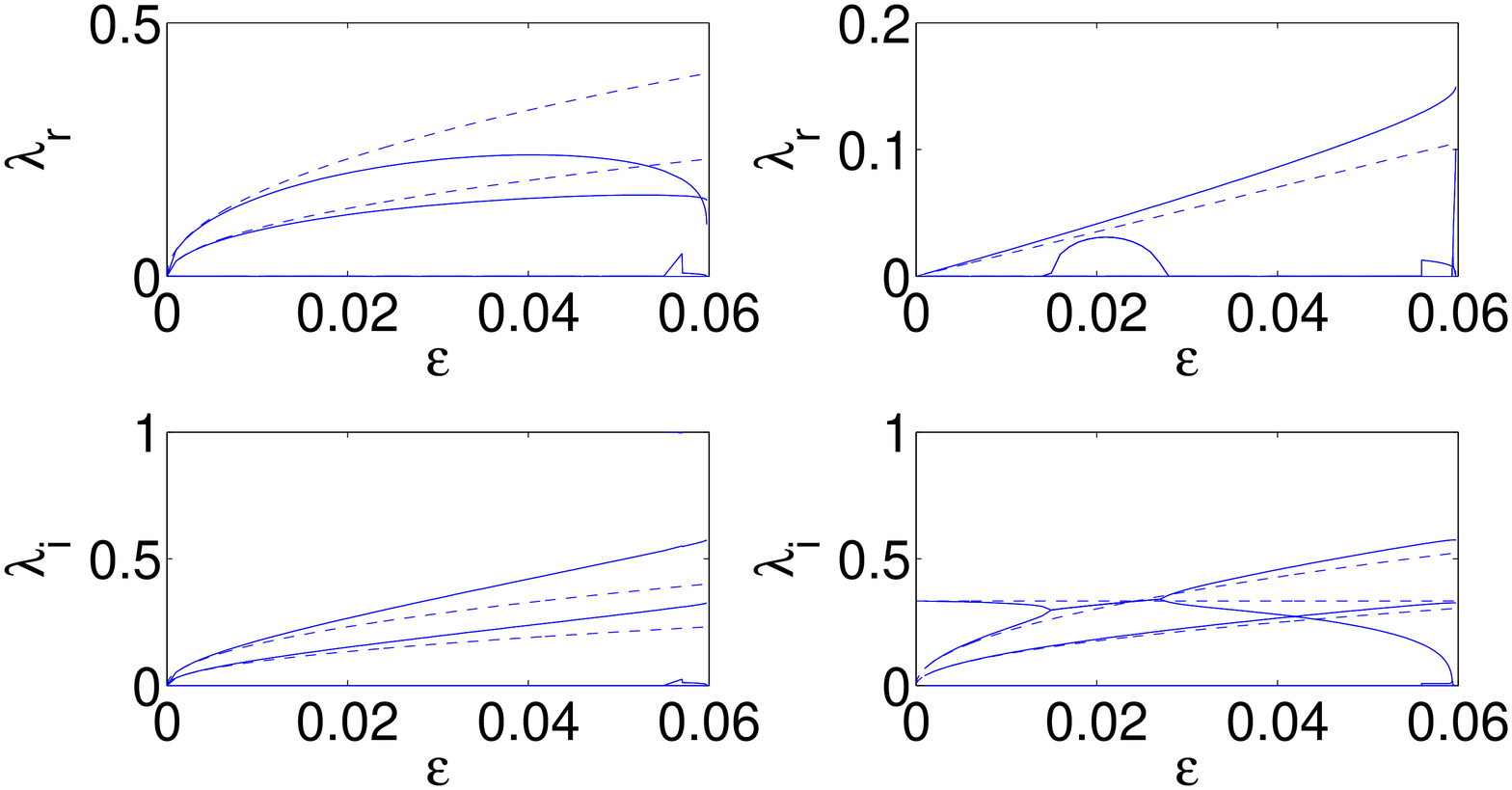}}
\caption{Similar as Fig.~\ref{fig+++,++-} but here  the left column
shows the mode +++,+-+,
which has eigenvalues $\lambda^2=\pm\frac{2}{\sqrt{5}}\epsilon,\
\pm\frac{6}{\sqrt{5}}\epsilon$ and the right column contains
the mode +0+,+-+, which
has eigenvalues $\lambda=\frac{1}{3}i,\ 1.24467\sqrt{\epsilon}i,\ 2.14026\sqrt{\epsilon}i,
\ 1.75819\epsilon$ (according to the theoretical analysis).
The first two pairs of eigenvalues collide and
form an eigenvalue quartet from $\epsilon=0.014$ to $0.028$.
The top four panels are captured when $\epsilon = 0.05$.}
\label{fig+++,+-+}
\end{figure}

\begin{figure}[htp]
\scalebox{0.375}{\includegraphics{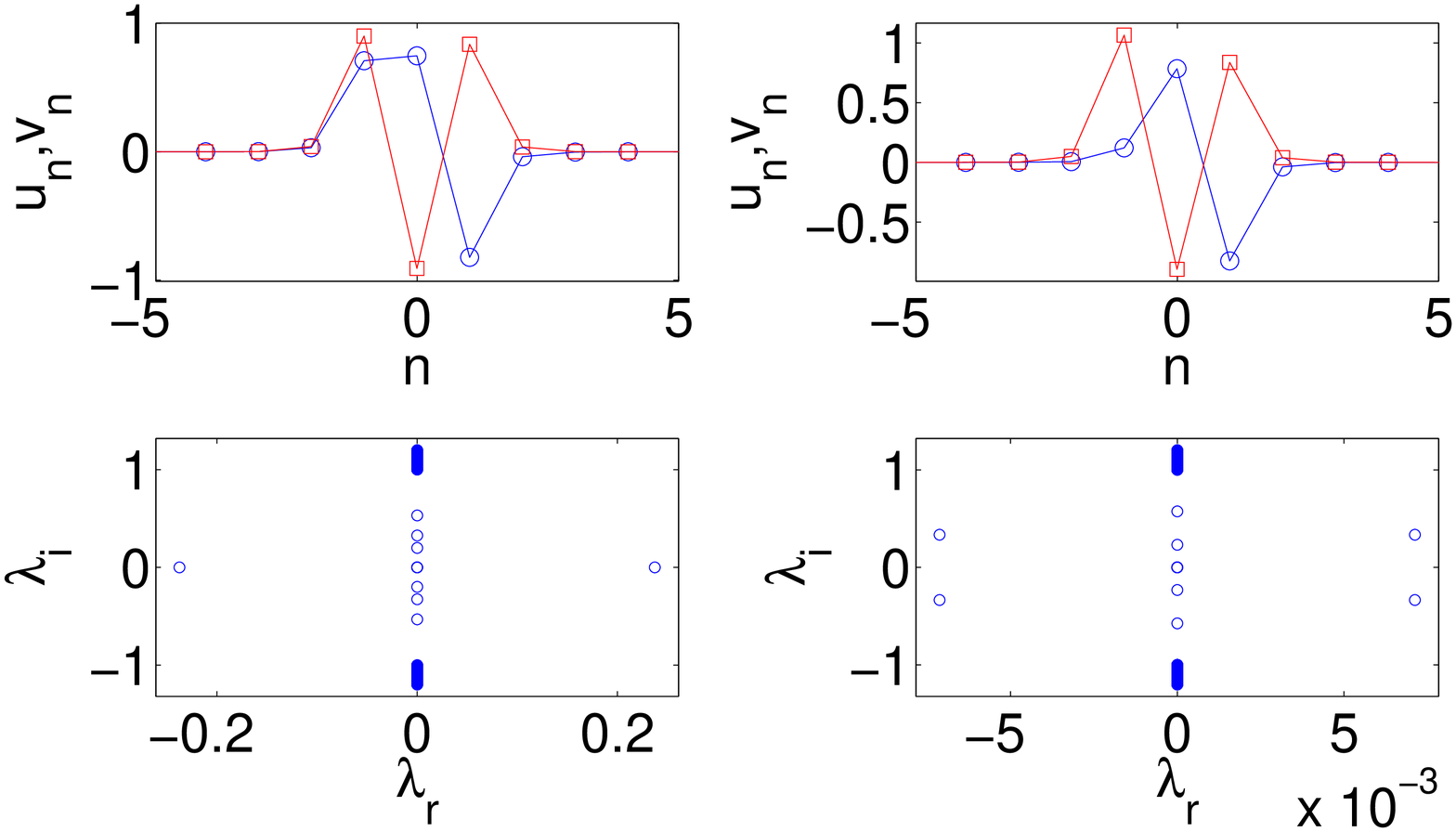}}
\scalebox{0.375}{\includegraphics{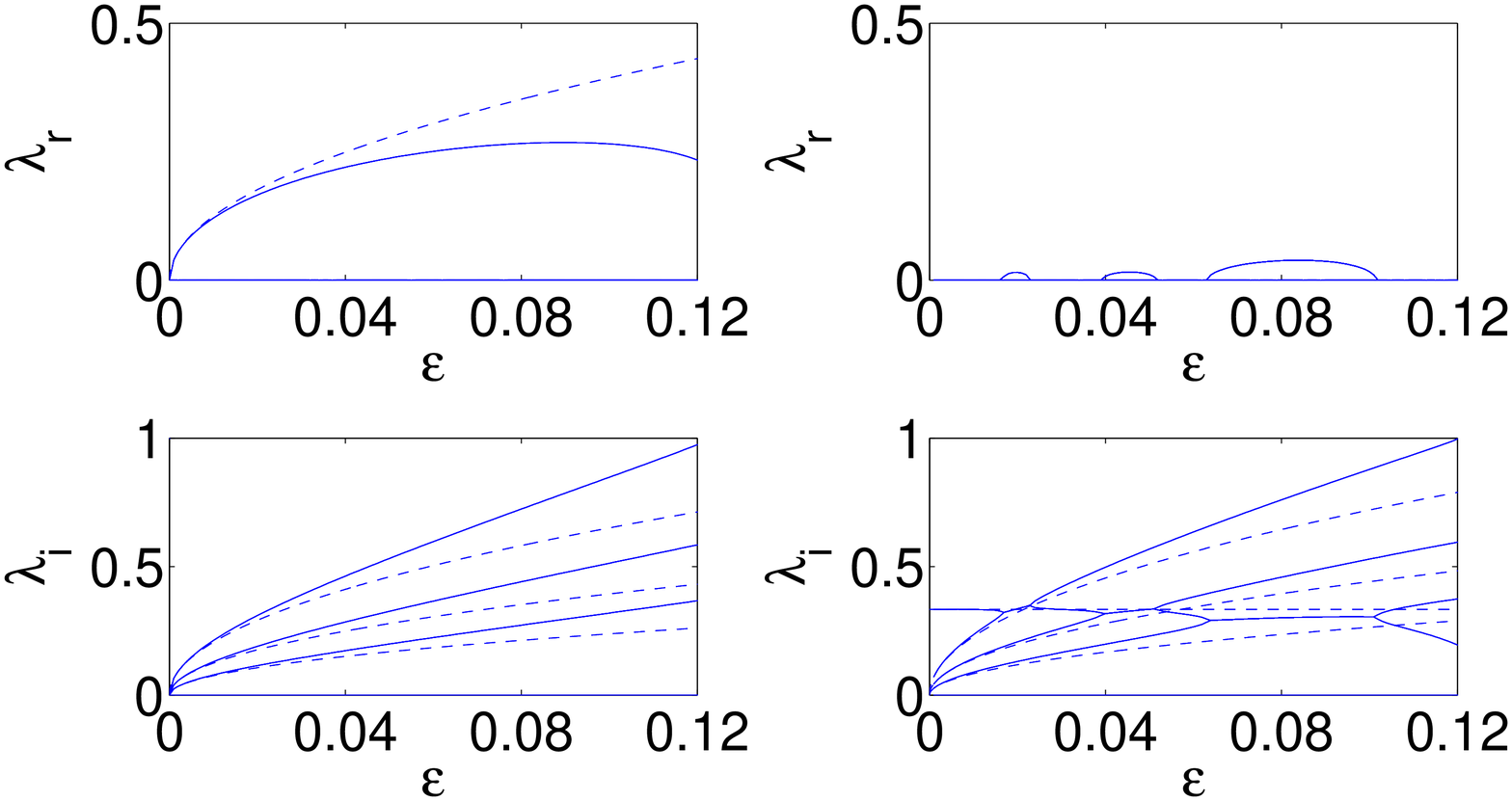}}
\caption{Similar as Fig.~\ref{fig+++,++-} but here the left column has
the mode ++-,+-+,
which has eigenvalues $\lambda^2=-\frac{2}{5} \left(6\pm\sqrt{21}\right)\epsilon,
\ \pm2 \sqrt{\frac{3}{5}}\epsilon$ while the right column contains
the mode 0+-,+-+, which has four
pairs of purely imaginary eigenvalues given by
$\lambda=\frac{1}{3}i,\ 0.834064\sqrt{\epsilon}i,\ 1.39576\sqrt{\epsilon}i,\ 2.27766\sqrt{\epsilon}i$.
The eigenvalues of order $\epsilon^0$ collide
with the others from $\epsilon=0.016$ to $0.023$, from
$0.039$ to $0.052$, and from $0.063$ to $0.102$, respectively, giving
rise to the corresponding oscillatory instabilities.
The top four panels are captured when $\epsilon = 0.05$.}
\label{fig++-,+-+}
\end{figure}

\begin{figure}[htp]
\scalebox{0.375}{\includegraphics{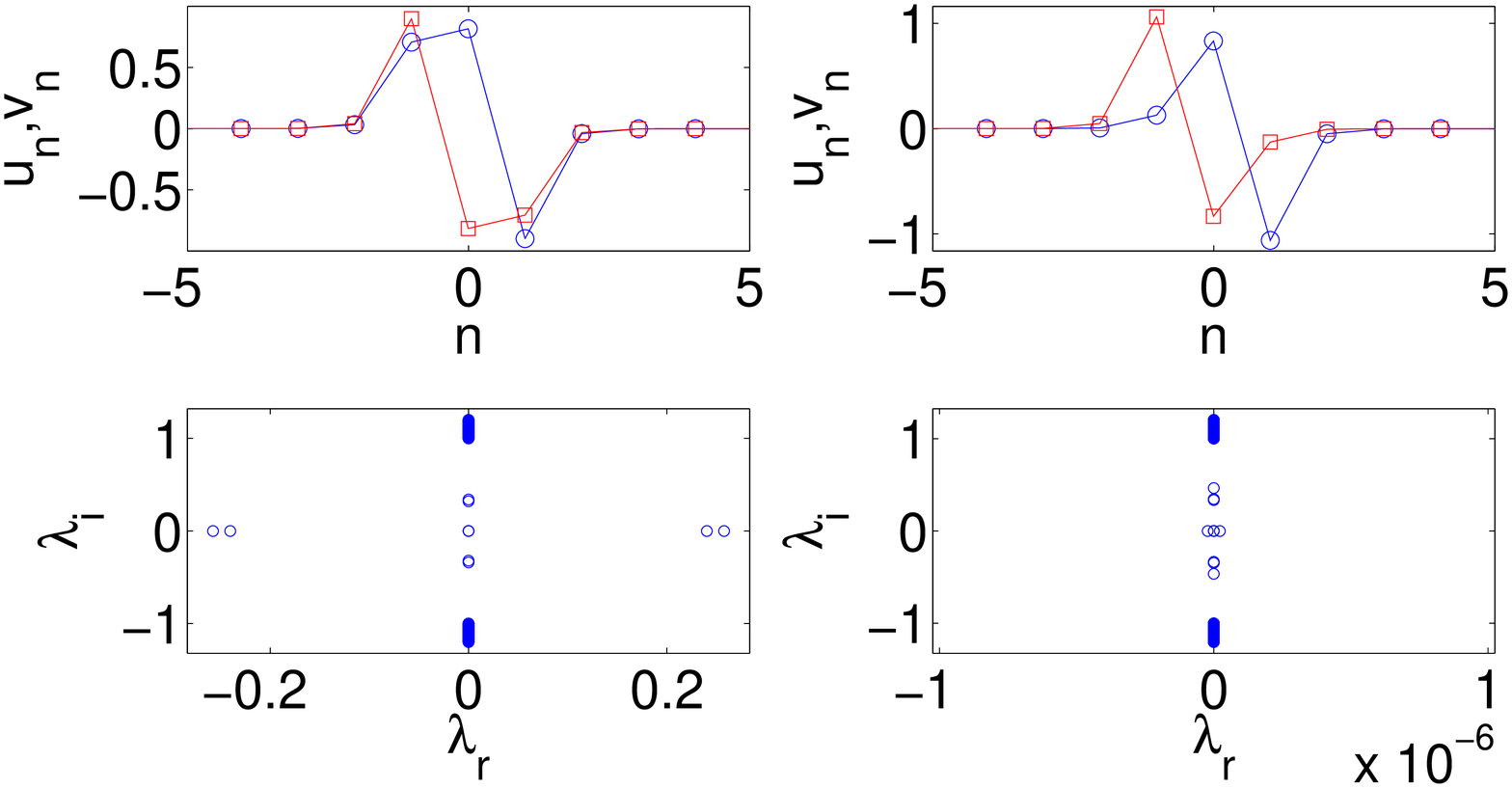}}
\scalebox{0.375}{\includegraphics{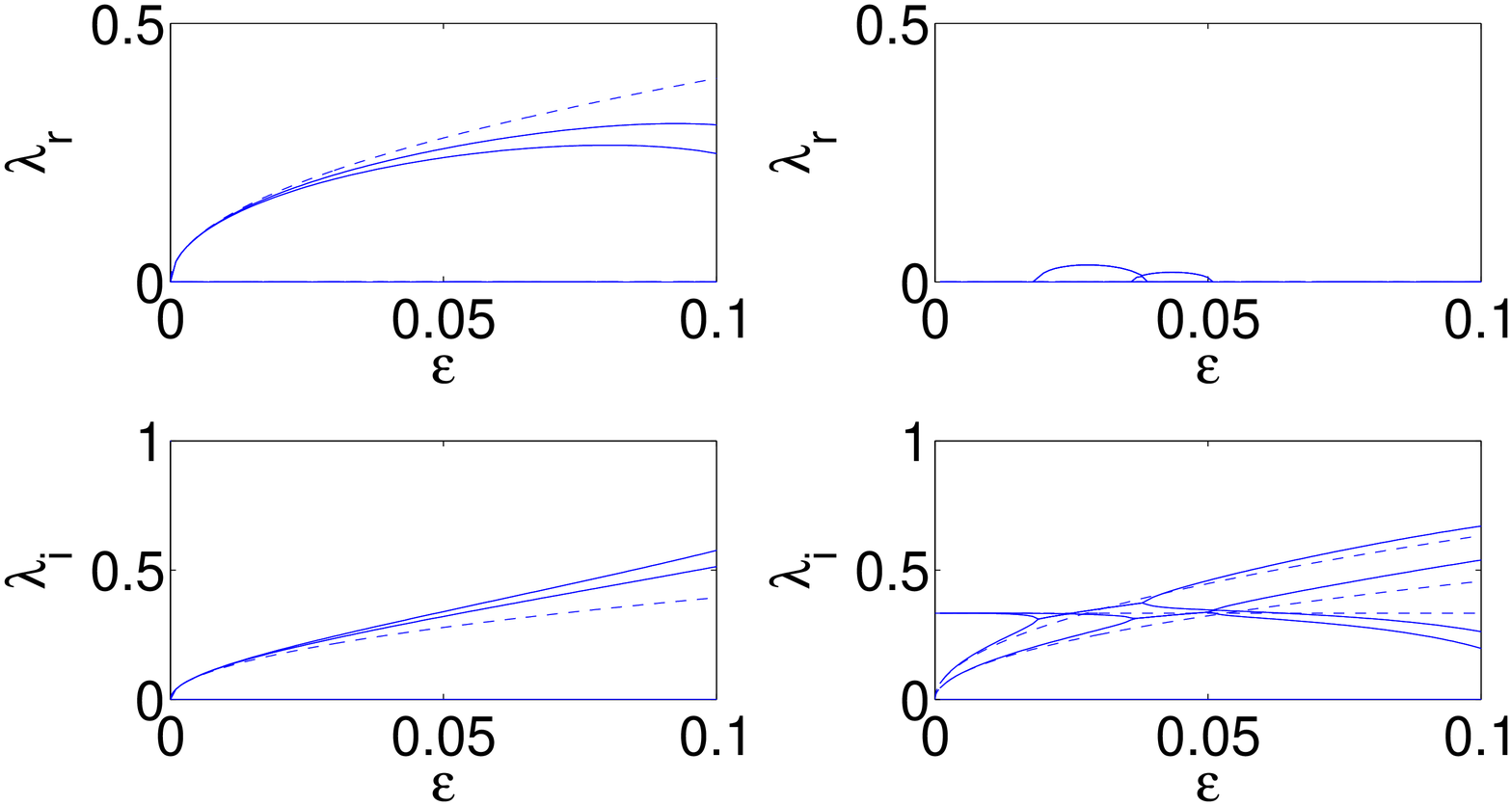}}
\caption{Similar as Fig.~\ref{fig+++,++-} but  the left column shows the mode
++-,+--,
which has four pairs of eigenvalues
$\lambda^2=\pm2\sqrt{\frac{3}{5}\epsilon},\
\pm2\sqrt{\frac{3}{5}\epsilon}$, while
the right column contains the mode 0+-,+-0, which has four
pairs of purely imaginary eigenvalues given by
$\lambda=\frac{1}{3}i,\ \frac{1}{3}i,\ 1.44578\sqrt{\epsilon}i,\ 2.00338\sqrt{\epsilon}i$.
The eigenvalues of order $\epsilon^0$ collide with the others
from $\epsilon=0.018$ to $0.039$, from
$0.036$ to $0.051$, respectively, creating once again an oscillatory
instability. The top four panels are captured when $\epsilon = 0.05$.}
\label{fig++-,+--}
\end{figure}

We now turn to a number of special case examples, which allow us to
consider the stability of the different two-site and three-site
configurations.

\section{Case examples}
We now focus on the set $S$: $N = 2$ and
$N = 3$. Since the phases of the two components
assume the values
$\xi_n,\ \zeta_n=\{0,\pi\}$ for $n\in S=\{1,2,\cdots,N\}$, we use
the symbolism ``+'' for $0$, ``-'' for $\pi$
and ``0'' for the case where no excitation is present on a particular site.

When $N=2$, the eigenvalue problem can be explicitly expressed in
terms of $g_{11},\ g_{12}$ and $g_{22}$ in the eigenvalues of
Eq. (\ref{M1}). More specifically, we obtain the following results.
\begin{itemize}
\item ++,++: The nonzero eigenvalues of $\mathcal M^{(1)}$ are given explicitly as
\begin{eqnarray}
\gamma_1&=&\frac{4 (g_{11}-g_{12}) (g_{12}-g_{22})}{-g_{12}^2+g_{11} g_{22}} \\
\gamma_2&=&-4.
\end{eqnarray}
Therefore, the ++,++ mode has two real unstable eigenvalues
\begin{eqnarray}
\lambda_1&=&\sqrt{\frac{4 (g_{11}-g_{12}) (g_{12}-g_{22})}{g_{12}^2-g_{11} g_{22}}\epsilon} \\
\lambda_2&=&2\sqrt{\epsilon}
\end{eqnarray}
in the stability problem (\ref{Hamiltonian}) for small $\epsilon>0$. Fig~\ref{fig++,++} shows the result
under parameters $g_{11}=g_{22}=1$ and $g_{12}=2/3$, where the
agreement with theory is very good for $\epsilon<0.05$. We have
generally
observed this kind of agreement with the analytical results presented
herein for an interval of $\epsilon$ values in the vicinity of the
anti-continuum limit. As a final observation, associated with
this case, we should point out that although the relevant eigenvalue
expression has non-trivial zero crossings, the constraints for the
existence of a real solution mandate that the relevant eigenvalue
always stay real and hence the configuration be
unstable~\footnote{Recall that in the Hamiltonian system under
consideration here, when $\lambda$ is an eigenvalue, so are $-\lambda$,
$\lambda^{\star}$ and $-\lambda^{\star}$, hence the existence of a
real
eigenvalue pair is always associated with an exponential instability.}.

\item ++,+-: In this case, we can obtain that
\begin{eqnarray}
\lambda_{1,2}^2&=&\frac{2 (g_{11} g_{12}-g_{12} g_{22})}{g_{12}^2-g_{11} g_{22}} \nonumber \\
&\pm&\frac{\sqrt{4 (g_{11}-g_{12}) (g_{12}-g_{22}) \left(g_{12}^2-g_{11} g_{22}\right)+(g_{11} g_{12}-g_{12} g_{22})^2}}{g_{12}^2-g_{11} g_{22}}.
\end{eqnarray}
In this setting, an interesting possibility arises, which deviates
from what is known about the one-component cases. In particular,
the nature of the instability of such configurations
depends on the choice of $g_{11},\ g_{12}$ and $g_{22}$.
When $g_{11}=g_{22}$ and $g_{12}>g_{11}$,
there is a quartet of complex conjugate eigenvalues which gives rise
to an oscillatory instability, as soon as $\epsilon$ becomes nonzero,
shown in Fig~\ref{fig++,+-_g12}.
On the other hand, when $g_{11}=g_{22}$ and $g_{12}<g_{11}$, a pair of
real eigenvalues and a pair of purely imaginary eigenvalues are
generated
as can be seen in the case example of the continuation
shown in  Fig~\ref{fig++,+-}.

\item +-,+-: The mode has two purely imaginary eigenvalues
\begin{eqnarray}
\lambda_1&=&\sqrt{-\frac{4 (g_{11}-g_{12}) (g_{12}-g_{22})}{g_{12}^2-g_{11} g_{22}}\epsilon} \\
\lambda_2&=&2\sqrt{-\epsilon}.
\end{eqnarray}
This is once again mandated by the existence constraints considered
previously.
As a result, similarly to the one-component analog of this
out-of-phase
mode, in this two-component installment too, we find that this mode
is linearly stable in the vicinity of the anti-continuum limit. This
mode
is only destabilized for sufficiently large values of $\epsilon
\approx 0.147$, when one of its imaginary eigenvalues collides
with the continuous spectrum and leads to a complex eigenvalue quartet.
The relevant detailed results for the continuation of this branch
are shown in  Fig.~\ref{fig+-,+-}.

\item ++,+0 and +-,+0: When $U_n$ is an excited site but $V_n$ is non-excited for some $n\in S$, we have
\begin{eqnarray}
\left(
\begin{array}{c}
\displaystyle\phi_n^{(0)} \\
\displaystyle\psi_n^{(0)}
\end{array}
\right)=
\left(
\begin{array}{c}
\displaystyle\pm\frac{1}{\sqrt{g_{11}}} \\
0
\end{array}
\right),\ \textit{or}\
\left(
\begin{array}{c}
0 \\
\displaystyle\pm\frac{1}{\sqrt{g_{22}}}
\end{array}
\right),
\end{eqnarray}
so that $\mathcal L^{(0)}\ne0$, which induces a nonzero eigenvalue $\displaystyle\lambda=|\frac{g_{12}}{g_{11}}-1|i$ or
$\displaystyle|\frac{g_{12}}{g_{22}}-1|i$, respectively, in the
eigenvalue problem (\ref{substitution}) of order $\epsilon^0$.
This already provides the dominant behavior of the corresponding
eigenvalues
as can be observed in the eigenvalue plots of Figs.~\ref{fig++,+0}
and \ref{fig+-,+0}. In addition to this purely imaginary eigenvalue,
there
is a small eigenvalue bifurcating from the origin in both of these
configurations, which is well approximated (for $g_{11}=g_{22}=1$ and
$g_{12}=2/3$ in the
figures), by $\lambda^2=\pm 1.74543 \epsilon$. However, this higher
order eigenvalue plays a critical role in the stability of the
configuration,
as in the case of the ++,+0 configuration this eigenvalue exists as
real (i.e., with the + sign in the above expression), while in the
latter
case of +-,+0, it is imaginary. Thus, the former is unstable and
the latter is stable, at least for small values of $\epsilon$;
for larger $\epsilon$, as can be seen in Fig.~\ref{fig+-,+0},
a collision arises between the O$(\epsilon^0)$ near-constant
eigenvalue and the growing O$(\sqrt{\epsilon})$ eigenvalue
(between $\epsilon=0.025$ and $\epsilon=0.041$) which gives rise to a complex eigenvalue quartet and hence an oscillatory
instability.
\end{itemize}

When $N=3$, analytical expressions for the eigenvalues $\lambda$ are
too cumbersome to obtain/write explicitly,
however
the relevant eigenvalues can be computed (as a function of $\epsilon$)
by solving a low dimensional
eigenvalue problem for each particular set of $g_{ij}$.
Fig~\ref{fig+++,+++}-\ref{fig++-,+--}
summarize the results for the three modes under the parameters $g_{11}=g_{22}=1$ and $g_{12}=2/3$.

The case of the +++,+++ configuration is shown in
Fig.~\ref{fig+++,+++},
where we can again verify the relevant expectations: for each excited
site, we expect one pair of emerging eigenvalues, hence 6 pairs should
be associated with the 3 excited sites in each component [notice that
in the case of 2 excited sites, this amounted to 4 such pairs].
Out of these, 2 are always set at the origin due to the phase
invariance
of each of the involved components, hence 4 pairs remain as
non-vanishing [2 in the 2-excited site case]. The in-phase structure
of the configuration renders all of them real in this case
in good agreement with the predicted theoretical eigenvalues
(of $\lambda^2=6\epsilon,\  2\epsilon,\  \frac{6}{5}\epsilon,\
\frac{2}{5}\epsilon$ given by the dashed lines in the graph).

The other two configurations which are identical across
components   are the ++-,++- of Fig.~\ref{fig++-,++-} and
the +-+,+-+ of Fig.~\ref{fig+-+,+-+}. In direct analogy to
their one-component counterparts examined in~\cite{pelin1},
we find that the former of these two configurations bears
2 real pairs of eigenvalues and 2 imaginary ones, while the
latter is indeed the only 3-site configuration that enjoys
linear stability in the vicinity of the anti-continuum
limit, due to the out-of-phase nature of the neighboring sites therein.
In the case of the ++-,++- configuration, the eigenvalues are
analytically found to be: $\lambda^2=\pm2\sqrt{3}\epsilon,\
\pm\frac{2\sqrt{3}}{5}\epsilon$, while in the case of the +-+,+-+,
the $\lambda^2$'s assume the opposite value than they do for the
+++,+++ case (yielding 4 imaginary pairs).

Finally, we consider configurations which are distinct in their
3-site form between the two components. These are shown in
Figs.~\ref{fig+++,++-}-\ref{fig++-,+--}.
It should be noted that some of these configurations
cease to exist within the interval of $\epsilon$ (up to $0.15$)
probed herein. In particular, for instance, the mode +++,++-
of Fig.~\ref{fig+++,++-} collides and disappears with the branch
+0+,++-, as can be seen in the figure; nevertheless in its
interval of existence, we can obtain good agreement with the
theoretical prediction for the corresponding eigenvalues
$\lambda^2=\frac{2}{5} \left(6\pm\sqrt{21}\right)\epsilon,\ \pm2
\sqrt{\frac{3}{5}}\epsilon$. In turn, for the mode +++,+-+,
we find that it collides with +0+,+-+ around $\epsilon=0.06$;
the details of the corresponding eigenvalues are shown in
Fig.~\ref{fig+++,+-+}. Lastly, the mode ++-,+-+ collides
with 0+-,+-+, as shown in Fig.~\ref{fig++-,+-+}, and
the mode ++-,+-- collides and disappears with 0+-,+-0,
as detailed in Fig.~\ref{fig++-,+--}.


\section{Conclusions \& future challenges}

In the present work, we considered two-component settings
and provided a general framework for the existence and stability
of localized mode solutions to nonlinear dynamical lattices
of the DNLS type. We illustrated that  $0$ and
$\pi$ relative phase configurations arise in this context,
in analogy to the case of the one-component DNLS model.
Although the stability presented some interesting deviations
from the corresponding one-component counterpart (e.g. in the
case of the ++,+- configuration), most of the cases presented
significant analogies in the nature of the stability conclusions
(with suitable modifications of the eigenvalue counts) to the
one-component setting. In particular, the existence of in-phase
near-neighbor excitations in the same component
appeared to generically lead to
the existence of real eigenvalue pairs in the linearization.
On the other hand, the only configurations that seemed to generically
be stable were those where all adjacent sites were out of phase with
respect to each other. In the case of the ++,+- configuration, the
unexpected deviation concerned the possibility that the form of
instability may not always
be of the kind that is encountered in the one-component case.
In particular, depending
on the relation between $g_{11}$, $g_{22}$ and $g_{12}$, the waveform
 may present either
exponential (as ++) or oscillatory (which +- would possess but far from the
AC limit) instabilities as soon as the coupling strength becomes nonzero.
This is a case where the inter-component
coupling affects the specific type of instability
observed (and can thus modify it in comparison to
the single component case).
On the other hand, more broadly, a global consideration of
the states analyzed herein suggests that we can
think of the stability (1) or instability (0)
of each component as a logical input. In that framework,
the stability of the two
component system operates in the form of an "AND" gate, and indeed
necessitates the stability of the relevant constituents.

It is natural to envision numerous generalizations of the present
considerations. While already some special case examples of
discrete vortex configurations have been examined in~\cite{pelin},
a general theory of both discrete solitons and of vortices is still
absent in the two-component setting. Far less work has been done
in the three-dimensional case, where identifying suitable stable
configurations
with nontrivial phase profiles is a fairly challenging task even
in the one-component case~\cite{pelin2}. While these considerations
are still lacking for the more standard DNLS model, this is certainly
all the more so the case for the setting with four-wave mixing terms
being present, as appears to be experimentally relevant per the work
of~\cite{christo}.
Finally, another possibility than the incoherent (phase-independent)
coupling presented herein or the four-wave mixing nonlinear coupling
mentioned above is the case where the two components sustain linear
coupling between them. In the latter case, even for single-site
excitations interesting phenomenologies have been illustrated e.g.
in~\cite{Herring2007}, such as symmetry breaking and pitchfork type
bifurcations. It would thus be especially relevant in that regard 
to examine multi-site modes and how their stability properties
will (potentially
dramatically due to the phase-coherent coupling) change in comparison
to the findings herein.
Some of these avenues are presently being
considered
and will be reported in future publications.

\vspace{5mm}

{\it Acknowledgments.} VR research has been co-financed by the European Union 
(European Social Fund – ESF) and Greek national funds through the Operational 
Program "Education and Lifelong Learning" of the National Strategic Reference 
Framework (NSRF) - Research Funding Program: THALES. Investing in knowledge 
society through the European Social Fund. PGK gratefully acknowledges
support from the National Science Foundation through grants
DMS-0806762 and CMMI-1000337, as well as from the Alexander von
Humboldt Foundation, the Alexander S. Onassis Public Benefit Foundation
and the Binational Science Foundation.

\bibliographystyle{elsart-num-sort}
\bibliography{biblio}

\begin{thebibliography}{10}
\expandafter\ifx\csname url\endcsname\relax
  \def\url#1{\texttt{#1}}\fi
\expandafter\ifx\csname urlprefix\endcsname\relax\def\urlprefix{URL }\fi

\bibitem{konotop1}
G.~Alfimov, V.~Brazhnyi, V.~Konotop, On classification of intrinsic localized
  modes for the discrete nonlinear schrodinger equation, Physica D 194 (2004)
  127.

\bibitem{general_review1}
S.~Aubry, Breathers in nonlinear lattices: Existence, linear stability and
  quantization, Physica D 103 (1997) 201.

\bibitem{konotop}
V.~Brazhnyi, V.~Konotop, Theory of nonlinear matter waves in optical lattices,
  Mod. Phys. Lett. B 18 (2004) 627.

\bibitem{zhig_two}
Z.~Chen, J.~Yang, A.~Bezryadina, I.~Makasyuk, Observation of two-dimensional
  lattice vector solitons, Opt. Lett. 29 (2004) 1656.

\bibitem{review_opt}
D.~N. Christodoulides, F.~Lederer, Y.~Silberberg, Discretizing light behaviour
  in linear and nonlinear waveguide lattices, Nature 424 (2003) 817--823.

\bibitem{interlaced}
J.~Cuevas, Q.~Hoq, H.~Susanto, P.~Kevrekidis, Interlaced solitons and vortices
  in coupled dnls lattices, Physica D 238 (2009) 2216.

\bibitem{detlef}
R.~Dong, C.~R{\"u}ter, D.~Kip, J.~Cuevas, P.~Kevrekidis, D.~Song, J.~Xu,
  Dark-bright gap solitons in coupled-mode one-dimensional saturable waveguide
  arrays, Phys. Rev. A 83 (2011) 063816.

\bibitem{efrem}
N.~K. Efremidis, S.~Sears, D.~N. Christodoulides, J.~W. Fleischer, M.~Segev,
  Discrete solitons in photorefractive optically induced photonic lattices,
  Phys. Rev. E 66 (2002) 046602.

\bibitem{7}
H.~Eisenberg, Y.~Silberberg, R.~Morandotti, A.~R. Boyd, J.~S. Aitchison,
  Discrete spatial optical solitons in waveguide arrays, Phys. Rev. Lett. 81
  (1998) 3383.

\bibitem{general_review3}
S.~Flach, A.~Gorbach, Discrete breathers - advances in theory and applications,
  Phys. Rep. 467 (2008) 1.

\bibitem{general_review2}
S.~Flach, C.~R. Willis, Discrete breathers, Phys.\ Rep. 295 (1998) 181.

\bibitem{rev_moti1}
J.~Fleischer, G.~Bartal, O.~Cohen, T.~Schwartz, O.~Manela, B.~Freedman,
  M.~Segev, H.~Buljan, N.~Efremidis, Spatial photonics in nonlinear waveguide
  arrays, Optics Express 13 (2005) 1780.

\bibitem{moti2}
J.~W. Fleischer, T.~Carmon, M.~Segev, N.~K. Efremidis, D.~N. Christodoulides,
  Observation of discrete solitons in optically induced real time waveguide
  arrays, Phys. Rev. Lett. 90 (2003) 023902.

\bibitem{moti1}
J.~W. Fleischer, M.~Segev, N.~K. Efremidis, D.~N. Christodoulides, Observation
  of two-dimensional discrete solitons in optically-induced nonlinear photonic
  lattices, Nature 422 (2003) 147.

\bibitem{Herring2007}
G.~Herring, P.~G. Kevrekidis, B.~A. Malomed, R.~Carretero-Gonz\'alez, D.~J.
  Frantzeskakis, Symmetry breaking in linearly coupled dynamical lattices,
  Phys. Rev. E 76 (2007) 066606.
\newline\urlprefix\url{http://link.aps.org/doi/10.1103/PhysRevE.76.066606}

\bibitem{hudock}
J.~Hudock, P.~Kevrekidis, B.~Malomed, D.~Christodoulides, Discrete vector
  solitons in two-dimensional nonlinear waveguide arrays: Solutions, stability,
  and dynamics, Phys. Rev. E 67 (2003) 056618.

\bibitem{pgk}
P.~Kevrekidis, The Discrete Nonlinear Schr¡§odinger Equation: Mathematical
  Analysis, Numerical Computations and Physical Perspectives, Springer-Verlag,
  Heidelberg, 2009.

\bibitem{ourbook}
P.~Kevrekidis, D.~Frantzeskakis, R.~Carretero-Gonz\'alez (eds.), Emergent
  Nonlinear Phenomena in Bose-Einstein Condensates: Theory and Experiment,
  Springer-Verlag, Heidelberg, 2008.

\bibitem{pelin}
P.~Kevrekidis, D.~Pelinovsky, Discrete vector on-site vortices, Proc. Roy. Soc.
  London A 462 (2006) 2671.

\bibitem{discrete_opt}
F.~Lederer, G.~Stegeman, D.~Christodoulides, G.~Assanto, M.~Segev,
  Y.~Silberberg, Discrete solitons in optics, Phys. Rep. 463 (2008) 1.

\bibitem{pelin2}
M.~Lukas, D.~Pelinovsky, P.~Kevrekidis, Lyapunov-schmidt reduction algorithm
  for three-dimensional discrete vortices, Physica D 237 (2008) 339.

\bibitem{christo}
J.~Meier, J.~Hudock, D.~Christodoulides, G.~Stegeman, Y.~Silberberg,
  R.~Morandotti, J.~Aitchison, Discrete vector solitons in kerr nonlinear
  waveguide arrays, Phys. Rev. Lett. 91 (2003) 143907.

\bibitem{us_dh}
K.~Mertes, J.~Merrill, R.~Carretero-Gonz{\'a}lez, D.~Frantzeskakis,
  P.~Kevrekidis, D.~Hall, Nonequilibrium dynamics and superfluid ring
  excitations in binary bose-einstein condensates, Phys. Rev. Lett. 99 (2007)
  190402.

\bibitem{KRb}
G.~Modugno, G.~Ferrari, G.~Roati, R.~Brecha, A.~Simoni, M.~Inguscio,
  Bose-einstein condensation of potassium atoms by sympathetic cooling, Science
  294 (2001) 1320.

\bibitem{morsch3}
O.~Morsch, E.~Arimondo, Dynamics and Thermodynamics of Systems with Long-Range
  Interactions, Springer, Berlin, 2002.

\bibitem{markus2}
O.~Morsch, M.~Oberthaler, Dynamics of bose-einstein condensates in optical
  lattices, Rev. Mod. Phys. 78 (2006) 179.

\bibitem{LiCs}
M.~Mudrich, S.~Kraft, K.~Singer, R.~Grimm, A.~Mosk, M.~Weidem{\"u}ller,
  Sympathetic cooling with two atomic species in an optical trap, Phys.\ Rev.\
  Lett.\ 88 (2002) 253001.

\bibitem{myatt}
C.~J. Myatt, E.~Burt, R.~Ghrist, E.~Cornell, C.~Wieman, Production of two
  overlapping bose-einstein condensates by sympathetic cooling, Phys. Rev.
  Lett. 78 (1997) 586.

\bibitem{pelin1}
D.~Pelinovsky, P.~Kevrekidis, D.~Frantzeskakis, Stability of discrete solitons
  in nonlinear schr{\"o}dinger lattices, Physica D 212 (2005) 1.

\bibitem{stamper}
D.~Stamper-Kurn, M.~Andrews, A.~Chikkatur, H.-J.~M. S.~Inouye, J.~Stenger,
  W.~Ketterle, Optical confinement of a bose-einstein condensate, Phys.\ Rev.\
  Lett.\ 80 (1998) 2027.

\end{thebibliography}

\end{document}